\documentclass{article}
\usepackage{arxiv}
\usepackage{cite}
\usepackage{amsmath,amssymb,amsfonts}
\usepackage{algorithmic}
\usepackage{graphicx}
\usepackage{multirow}
\usepackage{subfigure}
\usepackage{amssymb,amsfonts}
\usepackage{multicol}
\usepackage{textcomp}
\usepackage{outlines}
\usepackage{adjustbox,lipsum}
\usepackage{xspace}
\usepackage{xcolor}
\usepackage{wrapfig}
\usepackage{lscape}
\usepackage{wrapfig}

\definecolor{rev}{rgb}{0,0,0}
\newcommand{\rev}[1]{\textcolor{rev}{#1}}

\def\eg{\textit{e.g.}}

\def\etc{\textit{etc}}

\title{Outlier-based Autism Detection using \\Longitudinal Structural MRI\thanks{Accepted for publication in IEEE Access. Doi: https://doi.org/10.1109/ACCESS.2022.3157613}}
\author{Devika K, Venkata Ramana Murthy Oruganti \\
  Department of Electrical and Electronics Engineering, \\
  Amrita School of Engineering, Coimbatore, \\
  Amrita Vishwa Vidyapeetham, India \\
  \texttt{k\_devika@cb.students.amrita.edu, ovr\_murthy@cb.amrita.edu} \\
   \And
 Dwarikanath Mahapatra \\
  Inception Institute of Artificial Intelligence, \\
  Abu Dhabi, United Arab Emirates, \\
  \texttt{dmahapatra@gmail.com } \\
  \And
 Ramanathan Subramanian \\
  University of Canberra, \\
  BRUCE, ACT, Australia \\
  \texttt{ram.subramanian@canberra.edu.au} \\
}
\begin{document}
\maketitle
\begin{abstract} Diagnosis of Autism Spectrum Disorder (ASD) using clinical evaluation (cognitive tests) is challenging due to wide variations amongst individuals. Since no effective treatment exists, prompt and reliable ASD diagnosis can enable the effective preparation of treatment regimens. This paper proposes structural Magnetic Resonance Imaging (sMRI)-based ASD diagnosis via an outlier detection approach. To learn spatio-temporal patterns in structural brain connectivity, a Generative Adversarial Network (GAN) is trained exclusively with sMRI scans of healthy subjects. Given a stack of three adjacent slices as input, the GAN generator reconstructs the next three adjacent slices; the GAN discriminator then identifies ASD sMRI scan reconstructions as outliers. This model is compared against two other baselines-- a simpler UNet and a sophisticated Self-Attention GAN. Axial, Coronal and Sagittal sMRI slices from the multi-site ABIDE II dataset are used for evaluation. \rev{Extensive experiments reveal that our ASD detection framework performs comparably with the state-of-the-art with far fewer training data. Furthermore, longitudinal data (two scans per subject over time) achieve 17-28\% higher accuracy than cross-sectional data (one scan per subject). Among other findings, metrics employed for model training as well as reconstruction loss computation impact detection performance, and the coronal modality is found to best encode structural information for ASD detection.} 

\end{abstract}


\section{Introduction}
{Autism} Spectrum Disorder (ASD) is characterized as a developmental disability. Initial signs typically appear in the early stages of childhood~\cite{asd1, asd2}. Children with ASD are prone to a number of unusual or repetitive behavioral changes, including problems with speech, touch, eye contact and facial expression. These symptoms become more severe as age progresses~\cite{asd3, asd4, asd5,asd6, asd7}. Core ASD symptoms typically result due to developmental changes in structural and functional brain connectivities. The best way to detect and treat ASD is by effective diagnosis. Currently, the Diagnostic and Statistical Manual (DSM-5)~\cite{asd8, asd9}, Autism Diagnostic Observation Schedule (ADOS)~\cite{asd10, asd11} and the Autism Diagnostic Interview (ADI-R)~\cite{asd12, asd13} are used to conduct initial screening. Manual diagnoses are subjective, prone to errors and biases due to disparities in expertise~\cite{asd33, asd34}. Recently, Computer-Aided Diagnosis (CAD) is being used as an alternative aid in the diagnostic process~\cite{asd37, asd63}. 

Recent research has shown that neuroimaging analysis is useful for ASD diagnosis as Magnetic Resonance Imaging (MRI) methods are capable of examining both qualitative and quantitative details derived from detailed three-dimensional anatomical images~\cite{asd32}. Among imaging modalities, neuroimaging can be categorized into structural imaging and functional imaging. Structural MRI (sMRI) research relies on volumetric and morphometric studies to evaluate irregular neuroanatomy across the three acquisition planes -- Axial, Coronal and Sagittal. Functional MRI (fMRI), which utilizes the correlation between cerebral blood flow and brain activity, is preferred in studies which seek to examine how the nervous system oxidizes while undertaking visual, motor, and cognitive processes. In contrast, sMRI is popular in clinical studies~\cite{asd14, asd31} due to its ability to detect subtle brain structural changes, and to produce images with greater contrast and spatial resolution. 

The National Database for Autism Research (NDAR)~\cite{asd40b} and Autism Brain Imaging Data Exchange (ABIDE)~\cite{asd41a,asd41} are popular open-access databases for ASD research. Neuroimaging scans are obtained either as \textit{cross-sectional} (one sample per person) or \textit{longitudinal} (multiple samples per person captured over time) samples. These scans enable us to examine and monitor changes in brain structure and function in individuals over time. Most neuroimaging-based diagnostic research are based on cross-sectional data. Recently, a few researchers have analyzed longitudinal data to predict neurological disorders via machine learning~\cite{asd26}. Among the aforementioned databases, ABIDE II~\cite{asd41} provides longitudinal samples. Longitudinal data collection is difficult, as data needs to be acquired for the same subject at various time points. Subject readiness to engage in multiple scanning sessions is not assured in longitudinal setups, and subject-specific samples typically drop over time. A major drawback of longitudinal studies is therefore a limited sample size and fewer participants~\cite{asd39}.  

\rev{Conventional machine learning frameworks use various handcrafted features and classification techniques such as Support Vector Machine (SVM). However, the handcrafted features are the bottleneck for the success of the frameworks. In-depth domain knowledge and experience is usually required to design such handcrafted features. Differently, deep learning frameworks accomplish the same by intelligently learning the intricate bio-markers using substantial amount of training data. The learnt features, usually the outputs of initial layers of the deep neural nets, can sometimes be related to the handcrafted features diagnosed by the medical experts. Thus, deep learning frameworks complement, not replace the physician's regular diagnosis of medical disorders. ASD has been diagnosed via deep learning~\cite{asd72}, 
~\cite{asd62}.}
Among deep learning architectures, autoencoders~\cite{Goodfellow-et-al-2016} enable low-dimensional embedding of a high-dimensional input via an encoder--decoder block. We employ a Generative Adversarial Model (GAN)-based encoder-decoder framework for sMRI-based ASD detection. Learning structural brain connectivity, an encoder maps an sMRI image slice onto a low-dimensional vector; the decoder then reconstructs the \textit{next} slice from this embedding. The \textit{actual} and \textit{reconstructed} next slices are compared to compute the reconstruction loss, which is then back-propagated to train the GAN. When the GAN is exclusively trained with healthy sMRI scans, higher reconstruction losses would result for ASD scans due to structural connectivity differences between normal and ASD subjects. The GAN discriminator would therefore view ASD scans as \textit{outliers} (with reconstruction loss greater than threshold), enabling unsupervised ASD detection. 

Single slice reconstruction error~\cite{ asd16, asd17} has typically been employed as the objective for model training. Differently, we conjecture that structural connectives between  \textit{adjacent} sMRI slices capture class-specific characteristics better than single slices, and train the GAN model with stacks of three contiguous slices. We also evaluate three encoder-decoder architectures-- GAN~\cite{goodfellow2014generative}, UNet~\cite{journals/corr/RonnebergerFB15} and Self-attention GAN (SAGAN)~\cite{pmlr-v97-zhang19d} for detection efficacy. Further, most works on longitudinal ASD data analysis employ supervised learning where availability of sufficient ASD data is critical, but ASD data are scarce. Modeling ASD scans as outliers as in our approach addresses this issue. In summary, this paper makes the following research contributions. 

\begin{enumerate}
    \item We employ a GAN encoder-decoder framework for sMRI-based ASD detection. The GAN trained exclusively from healthy scans views ASD samples as outliers and enables ASD diagnosis. This approach obviates the need for many ASD training samples. 
    \item To effectively model structural brain connectives, stacks of three adjacent sMRI slices are input to the GAN to reconstruct the next three slices. Slice reconstruction loss is employed as the training objective. Empirical results (Table~\ref{Cross}) confirm that modeling structural patterns from three-slice stacks is more beneficial vis-\`a-vis single slices.
    \item We evaluate the GAN against a computationally less-intensive UNet, and more-intensive SAGAN. SAGAN outperforms the GAN and UNet architectures.
    \item We examine the efficacy of (a) longitudinal vs cross-sectional data (b) various loss functions for model training, and (b) the Axial, Coronal and Sagittal slices and their combinations for encoding structural information.  The L2$+$cosine loss objective is found to be the most effective, and a combination of the Axial and Coronal slices achieves the best detection performance.       
\end{enumerate}

The paper is structured as follows. A survey examining related work is presented in Section~\ref{RL}. Section~\ref{Meth} details our framework and other baselines. Section~\ref{ExRes} discusses empirical results, and the paper concludes in Section~\ref{Con}.

\section{Related work}\label{RL}
This section reviews longitudinal sMRI-based ASD diagnosis, and deep learning models developed to this end.

\subsection{Longitudinal Studies on ASD detection}

ASD detection has been attempted with both cross-sectional and longitudinal sMRI data. Wang \textit{et al.}~\cite{asd20} conducted a study of cerebellar thickness to determine longitudinal differences associated with ASD. The analysis used longitudinal scans from the ABIDE II  dataset, which includes 19 ASD subjects and 14 healthy subjects. Correlation between ADOS scores and lobular thickness data was examined. Subjects with ASD showed smaller lobular thickness and asymmetry in the right cerebellum, and this reduction is associated with the severity of behavioral symptoms. 

Fu \textit{et al.}~\cite{asd21} studied Gray Matter and White Matter association with respect to ASD using longitudinal sMRI and Diffusion Tensor Imaging for 34 ASD and 26 healthy subjects. Chi-square and $t$-tests were used to compare demographic and clinical features extracted from the baseline and follow-up scans for ASD and healthy subjects. The study discovered that at across time, Fractional Anisotropy (FA) and brain volume of white matter was higher in ASD subjects. Ning \textit{et al.}~\cite{asd45} analyzed the developmental patterns of core-symptom-anchored cortical vertex-wise Gyrification Index (GI) in ASD. They used data from 321 ASD and 350 healthy subjects from ABIDE I, and 14 ASD plus 7 healthy subjects' longitudinal data  from the ABIDE II dataset. Statistical differences  between two groups were examined using chi-square and $t$-tests. While comparing GI between the baseline and follow-up conditions, significant variations were discovered in ten ASD clusters, with nine clusters showing decreased gyrification. Prigge \textit{et al.}~\cite{asd46} reported longitudinal volumetric findings in ASD subjects acquired from FreeSurfer. Linear mixed-effects models were used to characterize longitudinal volumetric changes in the brain over time. ASD-specific findings included larger gray matter in early childhood, enlarged ventricles by early adulthood, and reduced corpus callosum volume in adulthood. Devika \textit{et al.}~\cite{asd22} investigated longitudinal sMRI samples from ABIDE II for supervised ASD detection, and reported a classification accuracy of 94.29\% using Support Vector Machines.

\subsection{Deep learning Models}
Mostafa \textit{et al.}~\cite{asd49} used Convolutional Autoencoder (CAE) for single sMRI slice reconstruction to diagnose ASD. The study used T1-weighted sMRI scans from 403 ASD and 468 healthy subjects from ABIDE-I~\cite{asd41a}. The CAE was trained with healthy subjects and tested with both ASD and healthy subjects. Similarity indices including Structural Similarity Index (SSIM), Mean squared Error (MSE), Peak Signal-to-Noise Ratio (PSNR) values were used as features for SVM and Linear Discriminant Analysis (LDA) classification. Baur~\textit{et al.}~\cite{asd57} developed an unsupervised  UNet model on MRI of healthy subjects to  detect variations corresponding to anomalous MRI. The model was tested on five different MRI datasets, and achieved a highest F1-score of 62\%. 


\subsection{Generative Models for Clinical Diagnosis}
\rev{In medical imaging, generative models have become popular for clinical diagnosis due to their ability to learn complex distributions from input samples~\cite{asd73, asd74}. Generative Adversarial Networks (GANs) and Variational Autoencoders (VAEs) are popular generative models, but GANs are advantageous as they do not explicitly compute probability densities and yield better results than VAEs via a game-theoretic approach. AnoGAN~\cite{asd76} was the first work to employ GAN to detect retinal anomalies using spectral-domain Optical Coherence Tomography (OCT). A Deep Convolutional GAN (DCGAN) was trained using 2D image patches compiled from clinical OCT volumes of healthy subjects. This model was tested with both healthy and pathological samples, and the weighted sum of the residual and discrimination losses was used to compute the anomaly score. Inspired by AnoGAN, unsupervised  metastatic bone tumor classification with GAN was proposed in~\cite{asd77}. Anomaly scores were determined by comparing the test image with a synthesized one at both the image and feature levels. Although AnoGAN demonstrated high performance, iterative techniques suffer from computing inefficiency for real-world applications, which was addressed via fast AnoGAN (f-AnoGAN)~\cite{asd78} that learnt a mapping from image to latent space with a Wasserstein GAN.}

\rev{Han \textit{et al.}~\cite{ad1} developed a two-step procedure for abnormality diagnosis from T1-weighted (T1w) sMRI Axial slices. Training was done on healthy subjects, while testing included both healthy and Alzheimer’s disease samples. UNet and GAN architectures were investigated, and a maximum AUC score of 0.92 was reported with the GAN architecture. Han \textit{et al.}~\cite{ad2} studied the effect of self-attention (SA) modules in GAN architectures to detect AD and brain metastases. The study used longitudinal samples, plus a cross-sectional dataset compiled by the authors, and reported a highest AUC score of 0.89 and 0.92 for Alzheimer's disease and brain metastases detection respectively. Translation from Abnormal-to-Normal GAN (ANT-GAN) is a variant of CycleGAN~\cite{asd79}, which was developed to synthesize a normal-looking medical image from an abnormal one, and vice versa. The proposed method was tested on MRI and CT scans from publicly available datasets. ANT-GAN was able to synthesize extremely realistic healthy scans that nearly matched images with lesions. Once a healthy scan is generated from an abnormal counterpart, discrepancies between the input and synthesized images can then be utilized to segment abnormal regions and contrast between healthy and abnormal scans. In ASD, GAN-generated synthetic data has significantly improved classification performance~\cite{asd75}}. 


\section{Methodology}\label{Meth}

\begin{figure*}[h]
\centering
\includegraphics[width=5.5in]{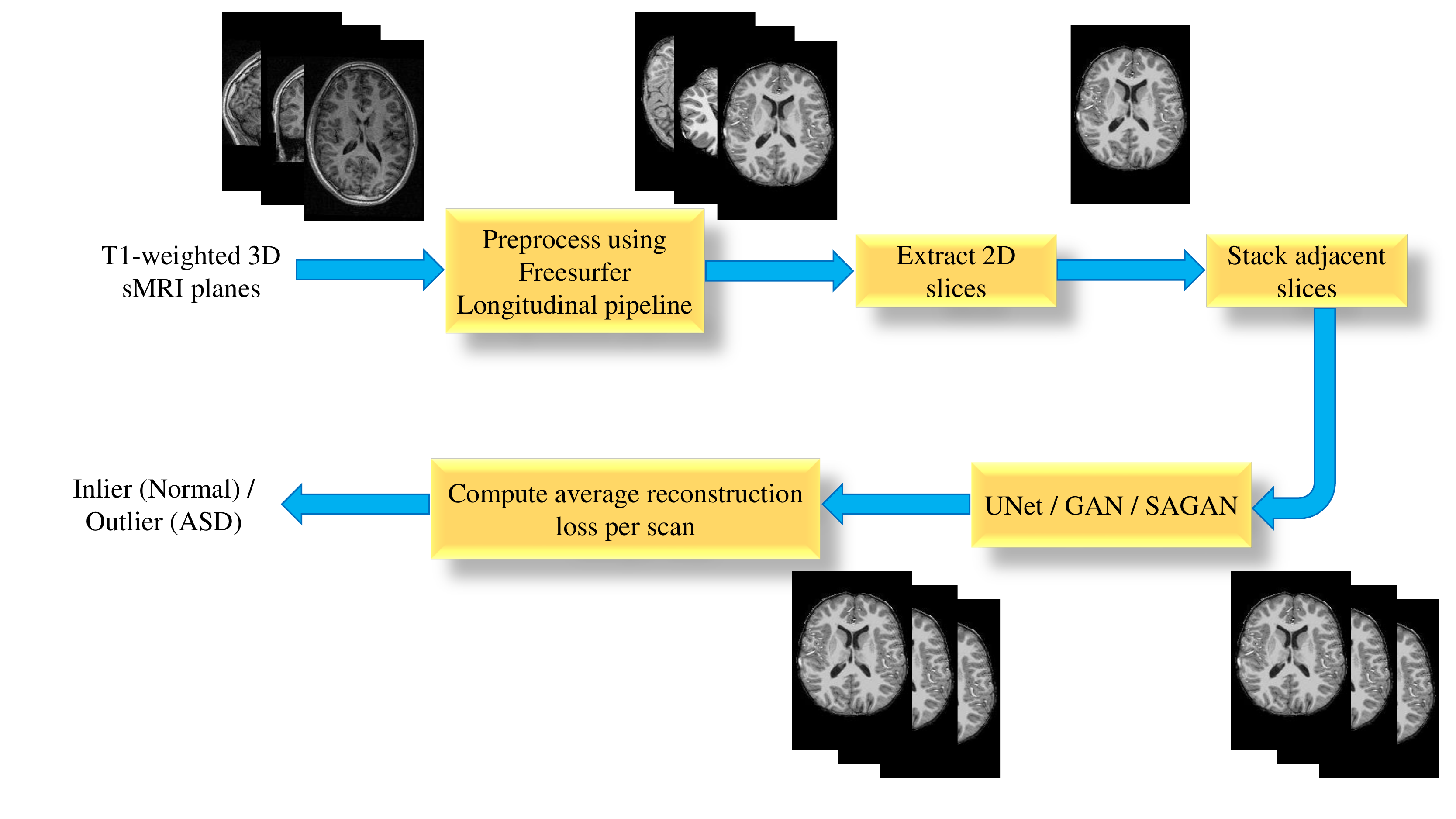}
\caption{\textcolor{black}{\textbf{Approach overview:} Longitudinal sMRI scans are pairwise-aligned via the Freesurfer pipeline. Slices along a given plane (Axial, Coronal and Sagittal) are extracted and three adjacent slices stacked. Different architectures (UNet, GAN, SAGAN) are trained using only sMRI scans of healthy subjects. Reconstruction error from training is used to compute a threshold for outlier direction. A test subject is labeled as ASD if reconstruction error is greater than the threshold.}} 
\label{fig1} 
\end{figure*}

An overview of the proposed framework is illustrated in Figure~\ref{fig1}. There are five stages. Firstly, we pre-process sMRI longitudinal scans via the Freesurfer longitudinal pipeline. Secondly, we extract 2D slices from longitudinally pre-processed 3D sMRI scans. Thirdly, Axial slices are selected for further processing. Fourthly, a GAN-based encoder-decoder framework is trained only using healthy subjects' data. The GAN objective is to reconstruct the next three adjacent slices from an input stack of (current) three adjacent slices. Finally, classification performance is evaluated based on the average loss between the reconstructed and ground-truth sMRI slices. Details of each stage and background are presented below.

\subsection{Dataset Description}
Autism Brain Imaging Data Exchange (ABIDE) is an open-source data collection with two subsets: ABIDE I~\cite{asd41a} was launched in 2014, and ABIDE II~\cite{asd41} in 2017.  ABIDE II includes data from 19 different sites, with longitudinal data from two sites (UCLA and UPSM)~\cite{asd41, asd42}. The scanner, scanning specifications, and scanning method are all different since each sMRI scan is gathered independently. \rev{In accordance with prior studies~\cite{asd49, asd52, asd69}, heterogeneity of the dataset is not explicitly addressed.  Longitudinal data from these two sites includes data for 23 ASD and 15 healthy subjects. The subjects ages range from 9--17 for baseline scan (median age of 12.6), and 10--19 at follow-up scan (median age of 15) across the healthy and ASD groups. The two longitudinal sets include T1-weighted (T1w) sMRI, rsfMRI scans in NIFTI format and phenotype information in comma separated value (.csv) format, collected during a one-to-two year period. By measuring the amount of water in the tissues, the sMRI data reveals the various types of tissues present. In T1w images, water and fluid-containing tissues look dark, while fat-containing tissues appear bright.}  This paper uses only longitudinal T1w sMRI slices for investigation.

\subsection{Data pre-processing}
Pre-processing is a necessary step for reducing inter-subject data variability resulting from data collection. Despite data being collected at different places, prior sMRI studies on ABIDE I and II only employed general pre-processing steps~\cite{asd14, asd52}. We followed the same suite and ignored inter-site scan capture variations. We employed the open-source Freesurfer (v.6.0) longitudinal pipeline to pre-process the T1w sMRI longitudinal samples~\cite{asd28, asd29}. Longitudinal processing requires cross-sectional processing followed by generation of a within-subject template (base image) via sequential, inverse-consistent registration of each time point scan to an average image. Following which, each time point scan is processed independently. The longitudinal pipeline is found to have higher cross-session dependencies than the cross-sectional pipeline~\cite{asd48}. We intend to leverage these dependencies with longitudinal samples. While pre-processing, three ASD subjects had surface reconstruction errors due to poor image quality, and hence their data were discarded.

\subsection{Multiple sMRI slice reconstruction}
\rev{Longitudinally pre-processed and skull-stripped 3D files from Freesurfer are fed to the Python3 utility \textit{med2image} (v.2.2.4). This package converts each slice (Axial, Coronal and Sagittal) of formatted 3D or 4D NIfTI (.nii) or DICOM (.dcm) medical image formats to common 2D image formats such as Joint Photographic Experts Group (jpg) or  Portable Network Graphics (png), suitable for training deep learning models~\cite{asd64, asd65}. We thus obtained slices along the Axial, Coronal and Sagittal planes. Among these, the Axial plane is popularly used~\cite{asd50, asd51}, but Convolutional Neural Networks (CNNs) have also effectively learned from Coronal plane~\cite{asd81}. In this study, we examined the utility of all three sMRI imaging planes.}  Of the 256 slices in a typical sMRI scan, slice sequence 120--180 is known to contain a majority of the vital brain information~\cite{asd49}. Hence, we extracted this 60-slice sMRI sequence for future stages. \rev{Extracted image dimensions ($ \text{height} \times \text{width}$) for the three planes are $256\times176$ (Axial), $256\times256$ (Coronal) and $256\times256$ (Sagittal). The Axial dimensions are specified by default in the later sections.} 

\begin{table}[h]
\caption{Details of train and test data used in our experiments.}
\begin{center} 
\centering
\begin{tabular}{c c} 
\hline
\textbf{Train} & \textbf{Test}  \\ 
\hline
\begin{tabular}[c]{@{}c@{}}\\\\ No. of healthy subjects = 12\\ No. of MRI scans = 24\\\\\\Total No. of samples = 24\\\end{tabular} & \begin{tabular}[c]{@{}c@{}}No. of healthy subjects = 3\\ Number of MRI scans = 6\\\\ No of ASD subjects = 20\\ Number of MRI scans = 40\\\\Total No. of samples = 46\\\end{tabular}  \\
\hline
\end{tabular}
\label{tab1}
\end{center}
\end{table}

Table~\ref{tab1} details the train and test splits for experiments conducted in this study. To ensure subject independence, we manually performed the train and test splits. Models were trained only with healthy subjects. The trained model is then tested with both normal and ASD samples, and unsupervisely detects ASD samples as outliers. 

We sought to model structural connectives between adjacent sMRI slices by feeding the GAN with, and reconstructing contiguous slices. To determine the optimum slice combinations, we considered the UNet model and conducted a preliminary experiment with four different input-output slice combinations, namely, 3-3, 3-5, 5-3 and 5-5. For instance, the UNet33 was trained with a 3-3 slice combination where the adjacent 3 slices (\eg, 1,2,3) of dimension $256\times176\times3$ were input to reconstruct the next adjacent 3 slices (\eg, 4,5,6) as in~\cite{ad1,ad2}. Sample input and predicted output slices~for the 3-3 and 3-5 combinations are shown in Table~\ref{tab11}. The same convention is used for the UNet35, UNet53, UNet55, GAN33 and SAGAN33 models. The L2 loss function was employed for training. The performance and computational time for each model on a computational cluster with a 28-core NVidia V100 GPU with 1.125TB RAM is shown in Table~\ref{tab12}. To maintain reasonable model training time and performance, we chose the 3-3 slice combination for our experiments.

\begin{table}[h]
\caption{Exemplar input and reconstructed slice sequences.}
\centering
\begin{tabular}{cclccl}
\hline
\multicolumn{2}{c}{\textbf{3-3}} &  & \multicolumn{2}{c}{\textbf{3-5}} \\ \hline
Input  & Reconstructed   &  & Input   & Reconstructed \\ Slices  &  Slices   &  & Slices     & Slices \\
\hline 
\hline
1,2,3   & 4,5,6   &  & 1,2,3   & 4,5,6,7,8 \\ \hline
2,3,4   & 5,6,7   &  & 2,3,4   & 5,6,7,8,9  \\ \hline
\multicolumn{2}{c}{\begin{tabular}[c]{@{}c@{}}..................\\ ......................\end{tabular}} &  & \multicolumn{2}{c}{\begin{tabular}[c]{@{}c@{}}..................\\ ....................\end{tabular}} \\ \hline
54,55,56  & 57,58,59    &  & 52,53,54   & 55,56,57,58,59  \\ \hline
55,56,57  & 58,59,60   &  & 53,54,,55   & 56,57,58,59,60  \\ \hline
\end{tabular}
\label{tab11}
\end{table}

\begin{table}[h]
\centering
\caption{UNet performance for different slice combinations.}
\begin{tabular}{cccc}
\hline
\textbf{S.No} & Slice & Accuracy & Training time \\
& combination & (\%) & (hrs) \\
\hline \hline
1& 3-3 (UNet33) & \textbf{36.95\% }  & 16   \\ \hline
2 & 3-5 (UNet35)  & 32.60\%    &\multirow{3}{*}{\begin{tabular}[c]{@{}c@{}}More than \\ 48 hours\end{tabular}} \\ \cline{1-3}
3   & 5-3 (UNet53) & 28.26\%   &    \\ \cline{1-3}
4    & 5-5 (UNet55) & 26.08\% &   \\ \hline
\end{tabular}
\label{tab12}
\end{table}

\subsection{Architecture details}
For slice reconstruction, we explored three networks, a GAN, a computationally less-intensive UNet and a more computationally intensive self-attention GAN (SAGAN). Their architectures are described below.

\subsubsection{UNet33 architecture}

\begin{figure*}[]
\centering
\includegraphics[width=0.95\linewidth]{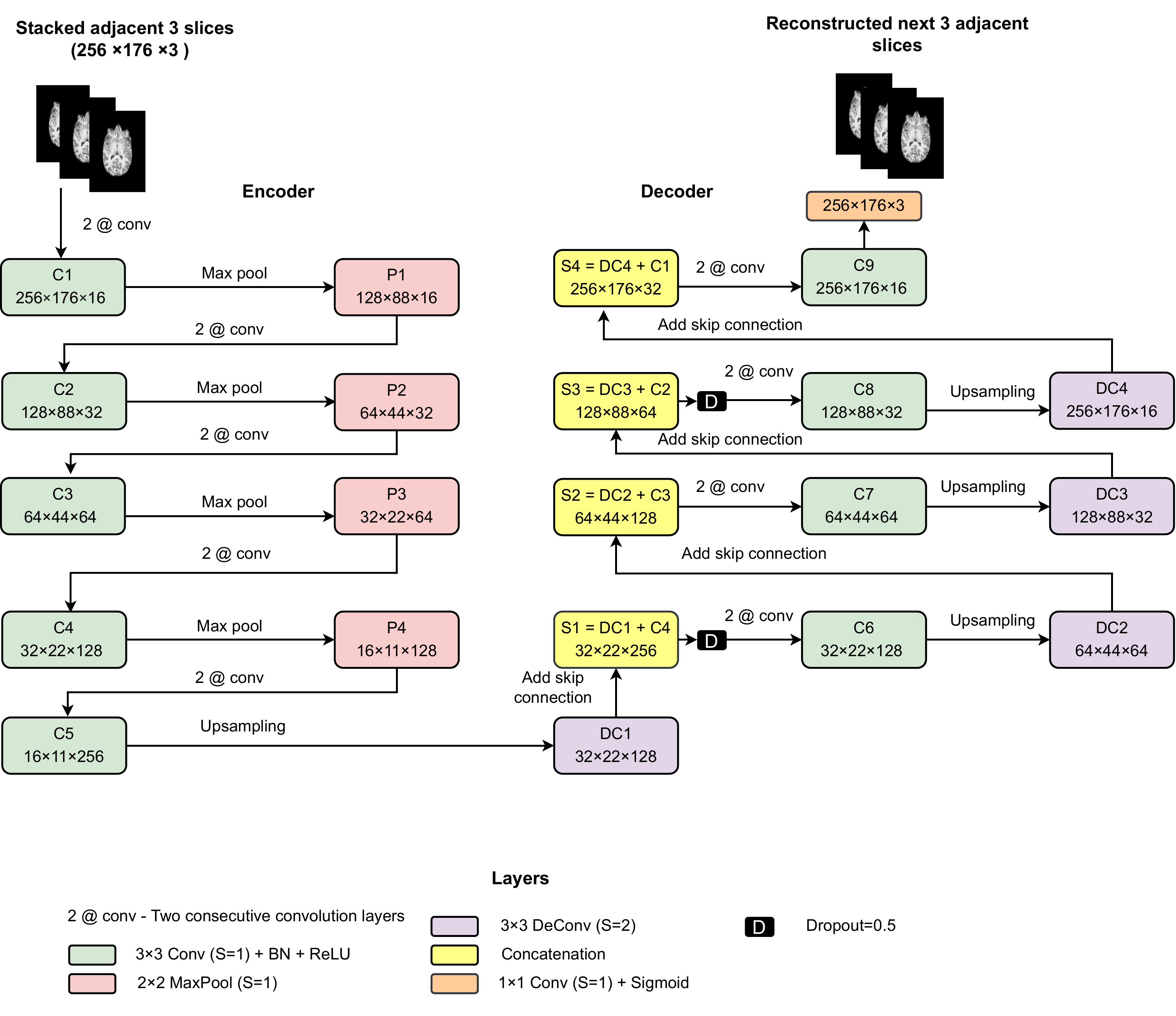}
\caption{Proposed UNet33 architecture includes nine normal and four transposed convolutional layers. Encoder receives an input slice of size $256\times176\times3$.\label{fig2}}
\end{figure*}  

The UNet33 architecture is illustrated in Figure~\ref{fig2}. There are two paths, namely, the \textit{contracting} path (encoder), and the \textit{expansion} path (decoder)~\cite{asd35}. The output dimensions ( $\text{height}\times \text{width} \times \text{channels}$) of each layer is specified within the boxes. Layers C1--C5 each denoting a series of two convolutional layers including batch normalization (BN) and rectified linear unit ReLU) activation, and max-pooling layers P1--P4 are part of the encoder, where the input image size and depth respectively decrease and increase from $256\times176\times3$ to $16\times11\times256$. Transposed convolutions (DC1, DC2, DC3, and DC4) are applied in the decoder, where the image size and depth respectively increase and decrease from $16\times11\times256$ to  $256\times176\times3$. Skip connections are added at all decoder stages (S1--S4) by concatenating the transposed convolution layer outputs with the corresponding encoder features. Every skip connection is followed by two regular convolutions (C6--C9), and a dropout of 0.5 (denoted by 'D') is performed at two points \rev{following skip connections to prevent overfitting~\cite{asd70, asd71}.} The output layer involves a $1\times1$ convolution with Sigmoid activation. Adam optimizer with a learning rate of  $2.0\times10\textsuperscript{-4}$, and the L2 loss function are employed for model training. 

\begin{figure*}[]	
\centering
\includegraphics[width=0.85\linewidth, height=15cm]{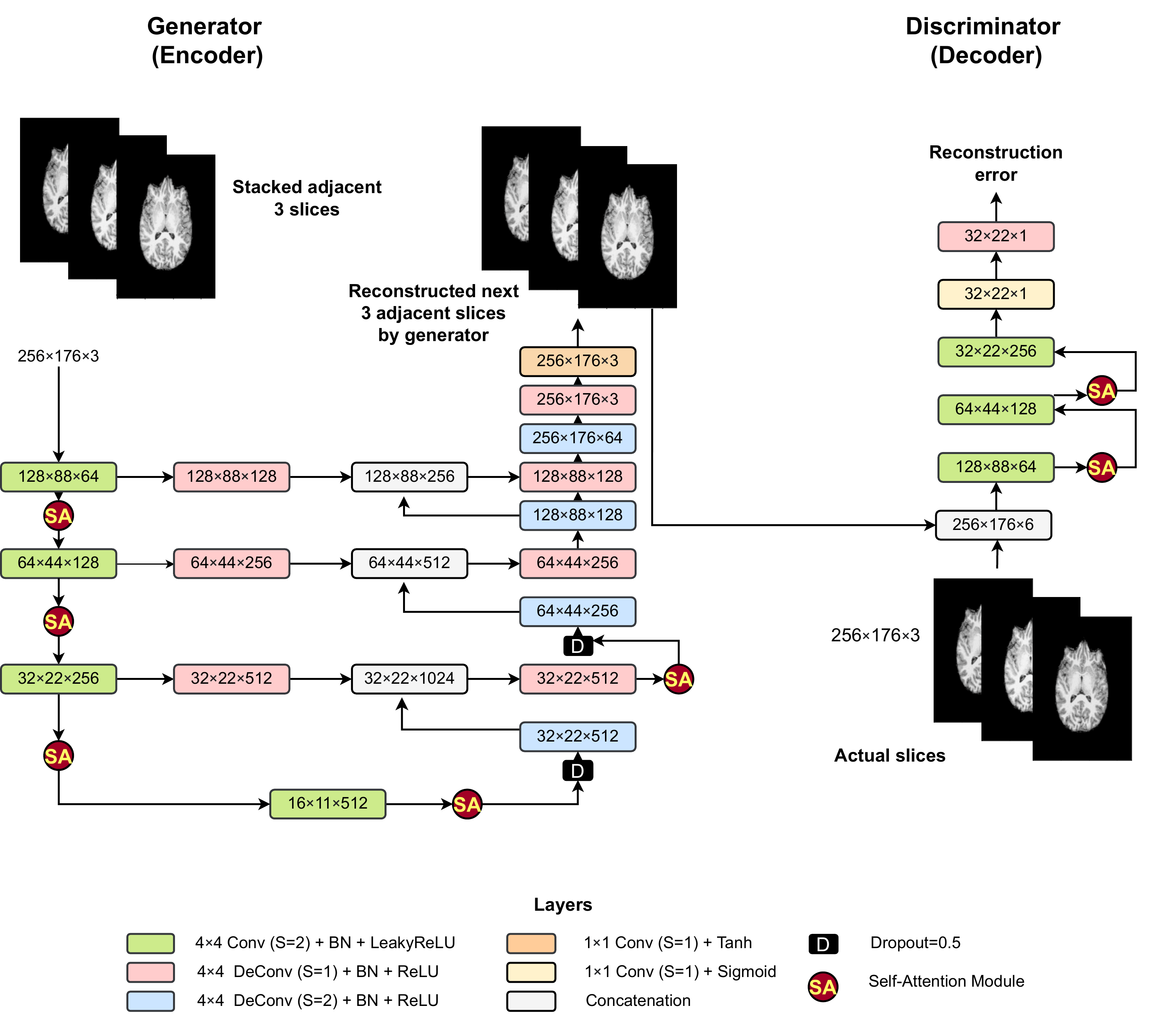}
\caption{Proposed GAN33 and SAGAN33 architecture for generating the next adjacent three slices from input adjacent three slices of size $256\times176$. GAN33 excludes SA modules, while SAGAN33 includes the seven SA modules.\label{fig3}}
\end{figure*} 

\begin{figure*}[!htbp]	
\centering
\includegraphics[width=0.8\linewidth, height=5cm]{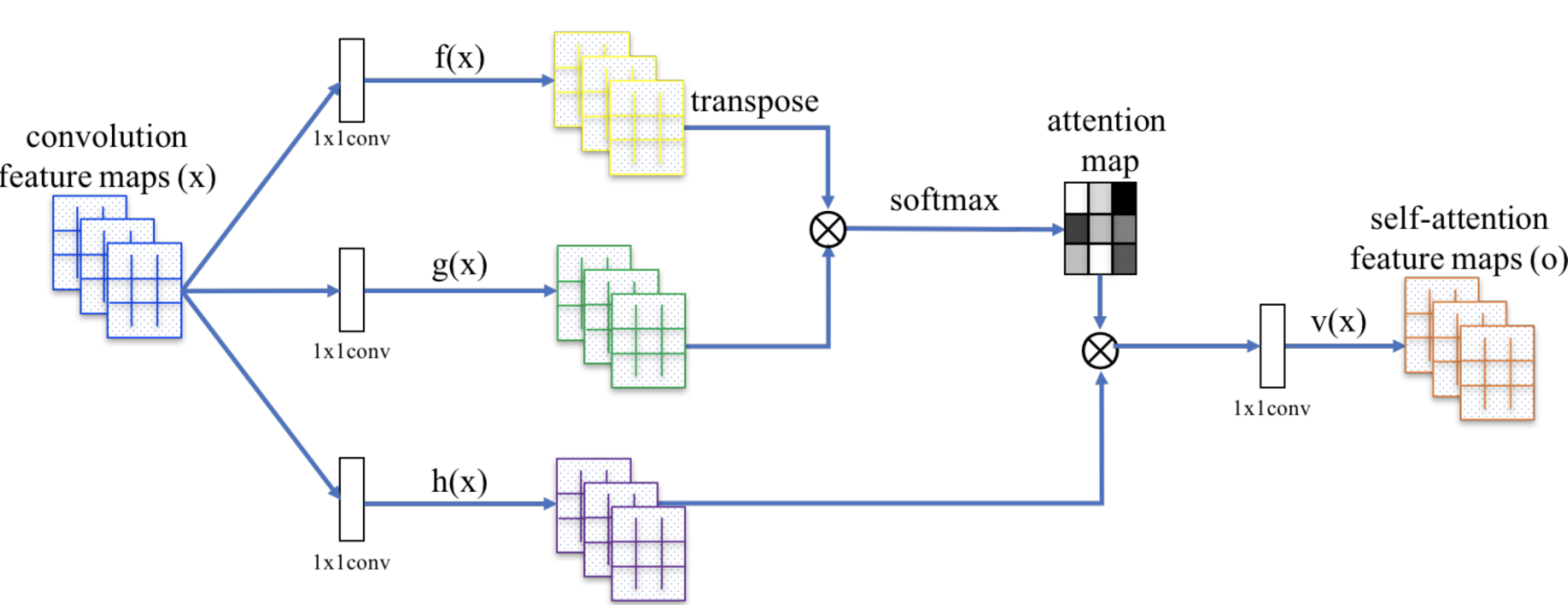}
\caption{\rev{Overview of the SA module~\cite{pmlr-v97-zhang19d} in the SAGAN33 architecture.}
\label {fig3a}}
\end{figure*}

L2 loss to measure reconstruction quality is defined as:
 \begin{equation}
 L2(X,Y) = \frac{1}{mn}\sum^{m-1}_{i=0}\sum^{n-1}_{j=0} [X_{i,j} - Y_{i,j}]^2
 \label{eq:1}
 \end{equation}
 
 \noindent where $X_{i,j}$ and $Y_{i,j}$ denote the $i^{th}$ ground truth/reconstructed slice stack of size 
 $n, i \in 1 \ldots m$. The input and reconstructed slice stacks are more similar as L2 loss decreases. We used early stopping with a patience threshold of ten epochs, and a batch size of 8. \rev{Hyperparameters were fine-tuned via grid search, and an input/reconstructed stack length of $m=3$ was fixed as it achieved the best accuracy as seen from Table~\ref{tab12}}.

\subsubsection{GAN33 architecture}
Figure~\ref{fig3} presents the architecture for GAN33. The GAN comprises two neural networks: a generator $G$ that receives input adjacent 3 slices and reconstructs the next adjacent 3 slices. The generator involves a UNet-like architecture with four $4\times4$ convolution layers in the encoder, and four $4\times4$ deconvolution layers (DeConv with stride = 2) in the decoder with same-level skip connections, and two dropout layers of 0.5. BN is applied to the convolutional and deconvolutional layers with Leaky ReLU and ReLU activation functions. The discriminator receives both the generated output and the ground-truth slice-stack. It uses 3 decoders. Given the training size, we used 1650 training steps with a batch size of 8, Adam optimizer with a learning rate of  $2\times10\textsuperscript{-4}$ and the WGAN-GP+100L1 loss function. The WGAN-GP is an advanced version of WGAN, and uses the gradient penalty for regularization; this increases training stability and prevents mode collapse~\cite{asd36}. We also employ the L1 loss as it facilitates a sharper reconstruction~\cite{asd23}. This WGAN-GP+100L1 loss function enables the synthesis of counterparts structurally similar to the ground-truth slices.

\subsubsection{SAGAN33 architecture}
SAGAN33 is a GAN33 with self-attention (SA) modules added as shown in Figure~\ref{fig3}. \rev{The SA module~\cite{pmlr-v97-zhang19d} is shown in Figure~\ref{fig3a}.
Three $1 \times 1$ convolutions are used to segregate feature maps acquired from the previous convolution layer. The SA mechanism is applied over feature maps obtained from the transformations \textit{f}, \textit{g} and \textit{h}.} This ensures that distant image parts are compatible with each other, unlike normal GANs~\cite{pmlr-v97-zhang19d}. Long-range dependencies among the image regions is established via the SA mechanism. The local and global image dependencies are combined to enhance details and quality of the reconstructed images. Seven SA modules are included in the SAGAN-- five SA modules in the generator, and two SA modules in the discriminator.  These layers are complementary to the convolutional layers, and allow the network to capture finer information. Output size of the SA modules is identical to the input~\cite{asd25}. Hyperparameters for SAGAN33 were set identical to GAN33.

\begin{figure*}%
\centering
\subfigure[Input]{%
\label{fig:1}%
\includegraphics[height=3in, width=0.15\linewidth ]{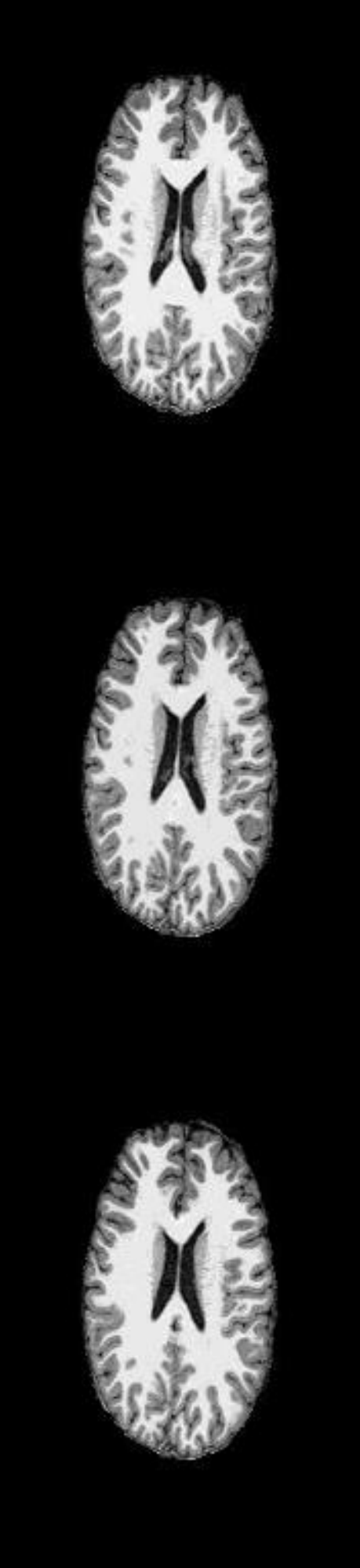}}%
\qquad
\subfigure[Ground truth]{%
\label{fig:2}%
\includegraphics[height=3in, width=0.15\linewidth]{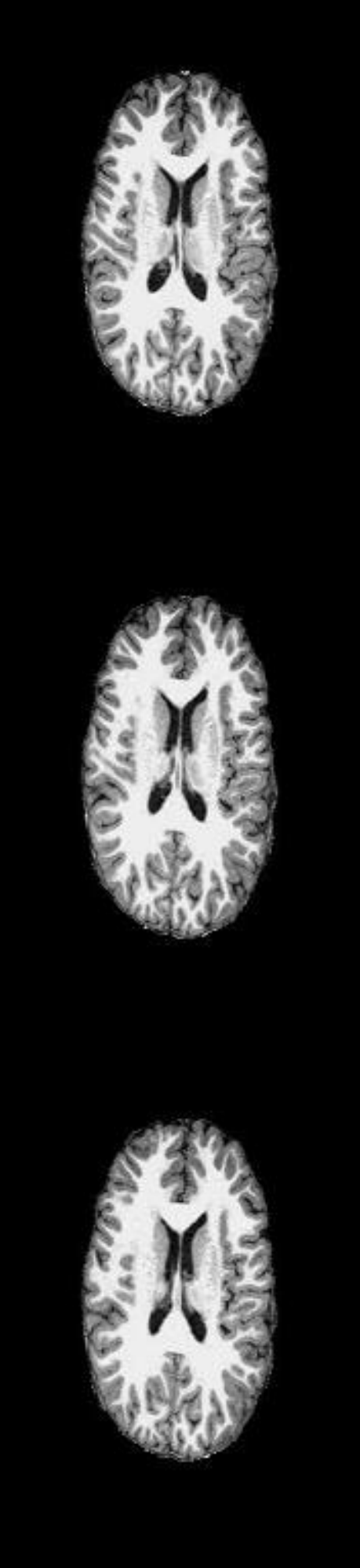}}%
\qquad
\subfigure[UNet33 (47.08)]{%
\label{fig:3}%
\includegraphics[height=3in, width=0.15\linewidth]{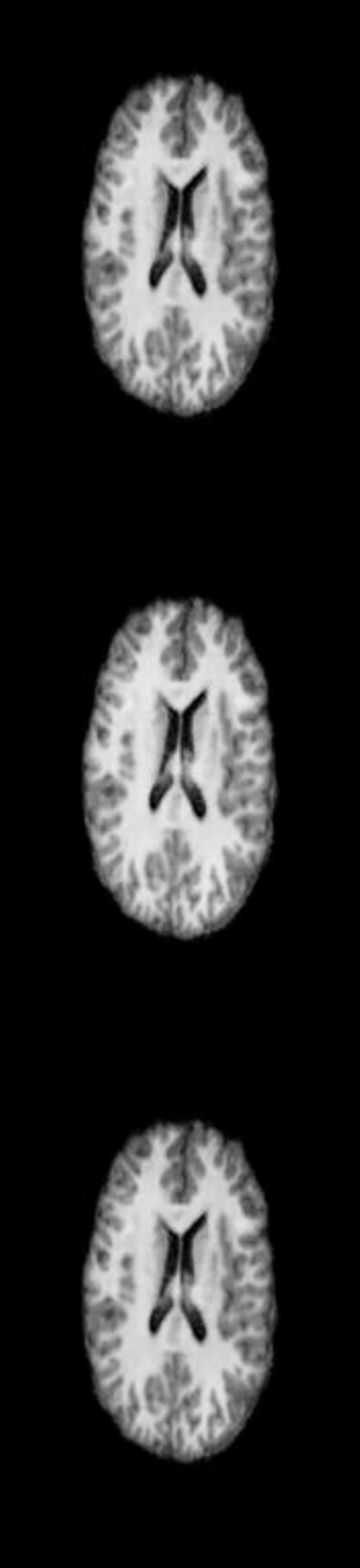}}%
\qquad
\subfigure[GAN33(47.52)]{%
\label{fig:4}%
\includegraphics[height=3in, width=0.15\linewidth]{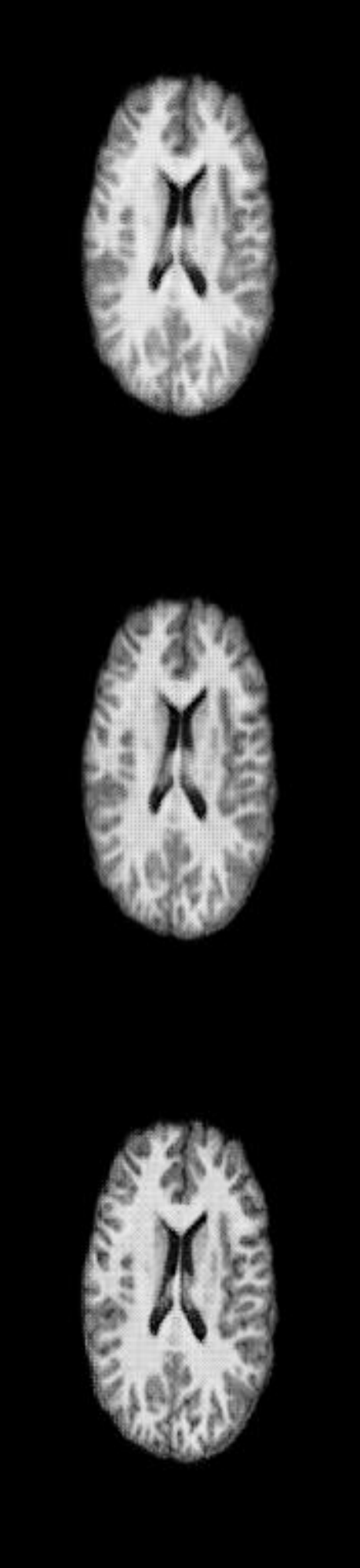}}%
\qquad
\subfigure[SAGAN33(47.61)]{%
\label{fig:5}%
\includegraphics[height=3in, width=0.15\linewidth]{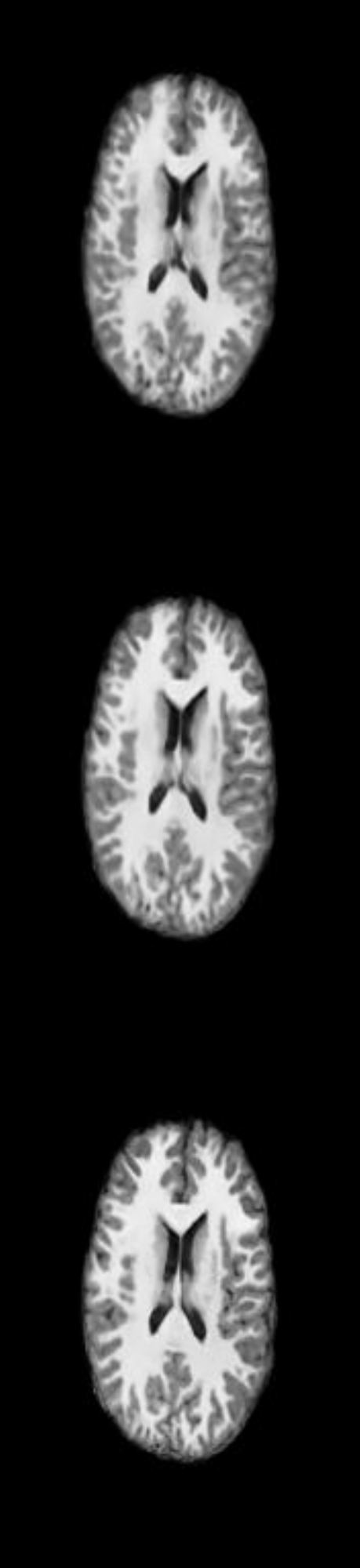}}%
\caption{Exemplar Axial slice reconstruction for a healthy sMRI sample: (a) Input adjacent-3 slices; (b) Original next adjacent three-slices; Reconstructed slices with (c) UNet33, (d) GAN33 and (e) SAGAN33. Reconstruction PSNR is specified in brackets.}
\label{fig4}
\end{figure*}

\begin{figure*}%
\centering
\subfigure[Input]{%
\label{fig:1a}%
\includegraphics[height=3in, width=0.15\linewidth ]{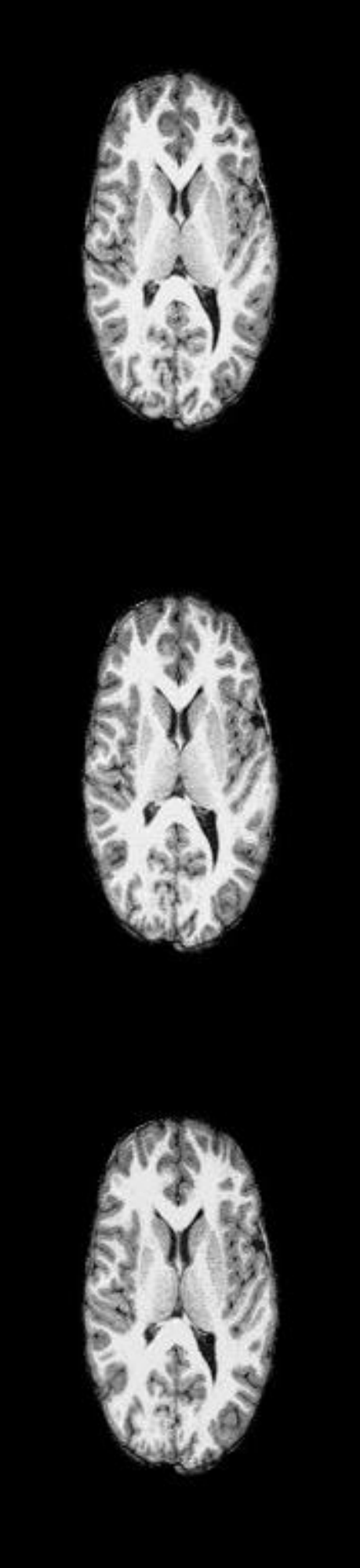}}%
\qquad
\subfigure[Ground truth]{%
\label{fig:2a}%
\includegraphics[height=3in, width=0.15\linewidth]{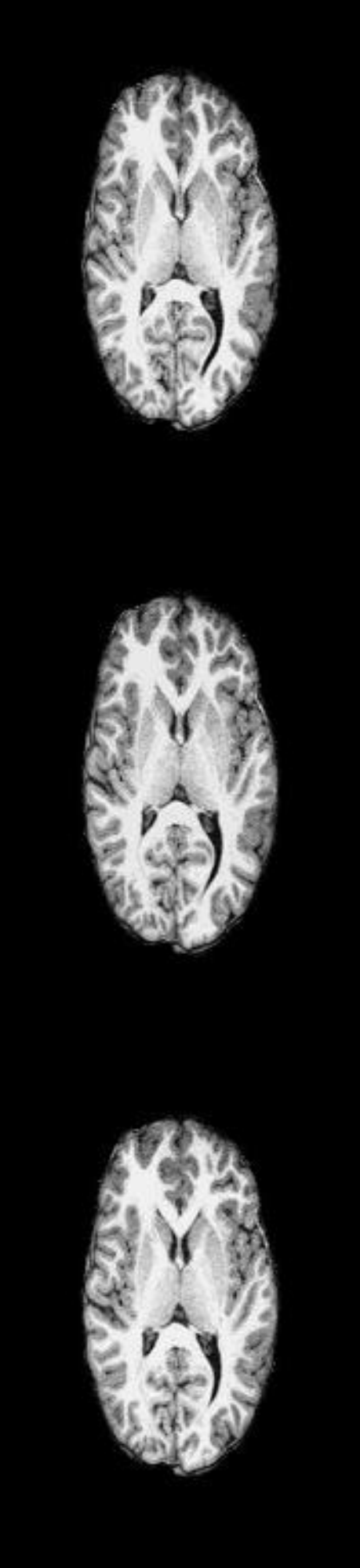}}%
\qquad
\subfigure[UNet33(46.46)]{%
\label{fig:3a}%
\includegraphics[height=3in, width=0.15\linewidth]{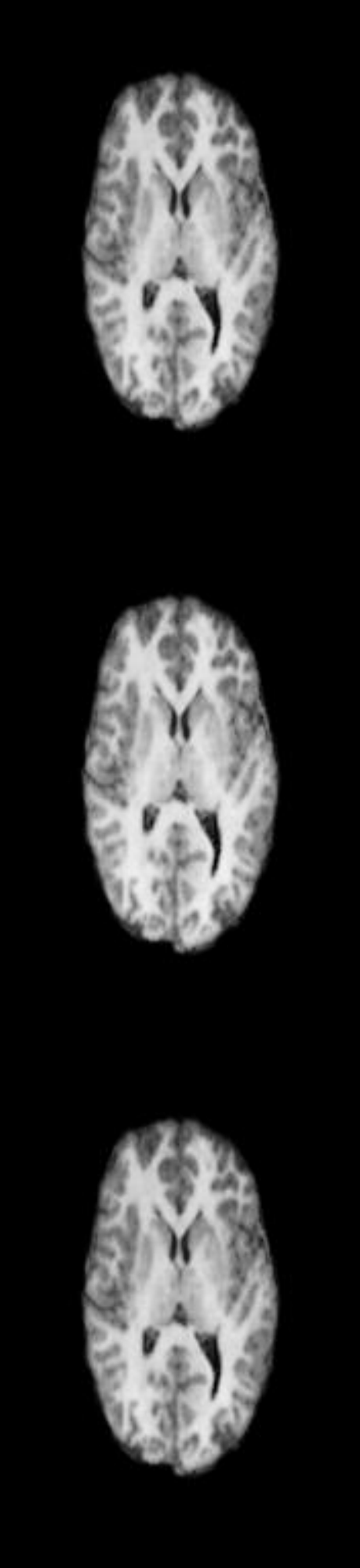}}%
\qquad
\subfigure[GAN33(46.90)]{%
\label{fig:4a}%
\includegraphics[height=3in, width=0.15\linewidth]{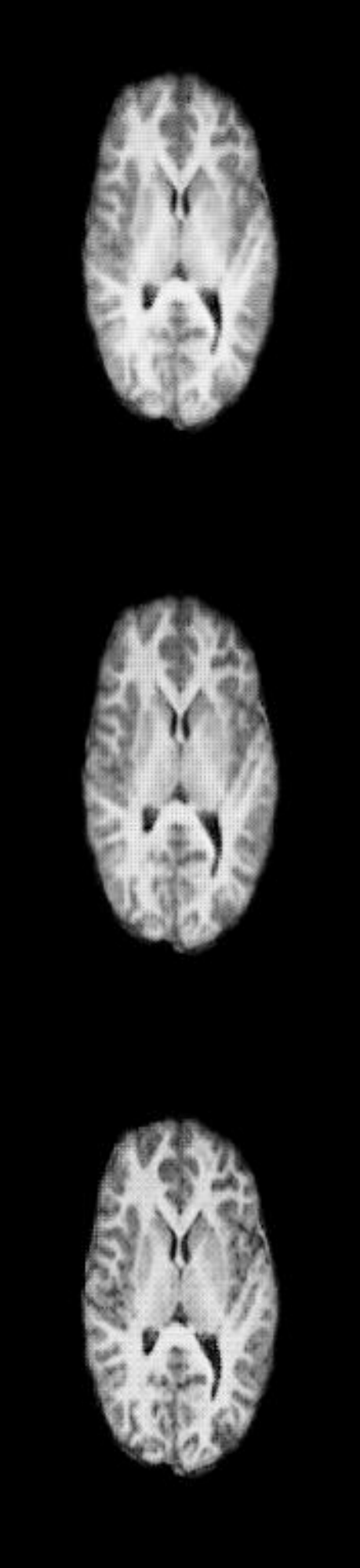}}%
\qquad
\subfigure[SAGAN33(47.03)]{%
\label{fig:5a}%
\includegraphics[height=3in, width=0.15\linewidth]{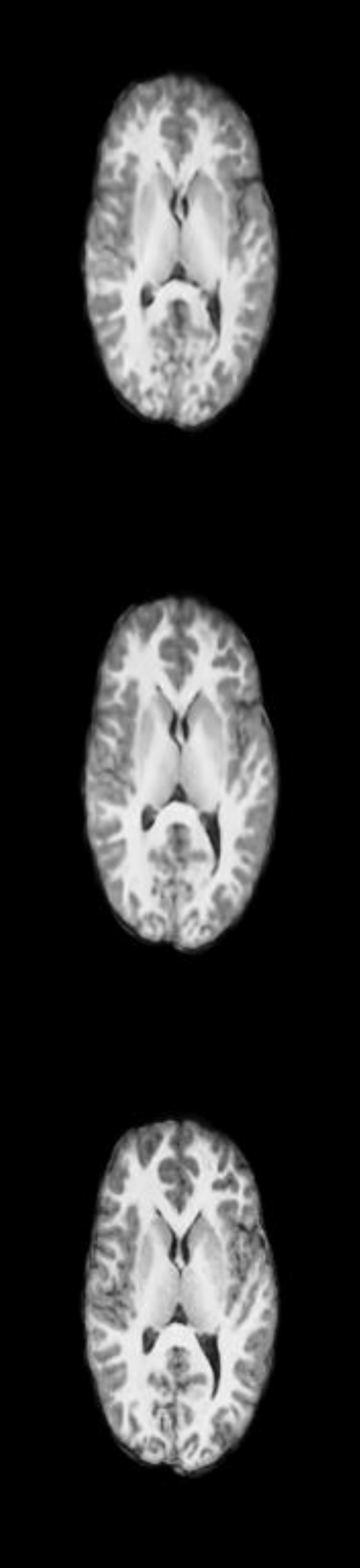}}
\caption{Exemplar Axial slice reconstruction for an ASD sMRI sample: (a) Input adjacent-3 slices; (b) Original next adjacent three-slices; Reconstructed slices with (c) UNet33, (d) GAN33 and (e) SAGAN33. Reconstruction PSNR is specified in brackets.}
\label{fig5}
\end{figure*}

\begin{figure*}%
\centering
\subfigure[Input]{%
\label{fig:1b}%
\includegraphics[height=3in, width=0.15\linewidth ]{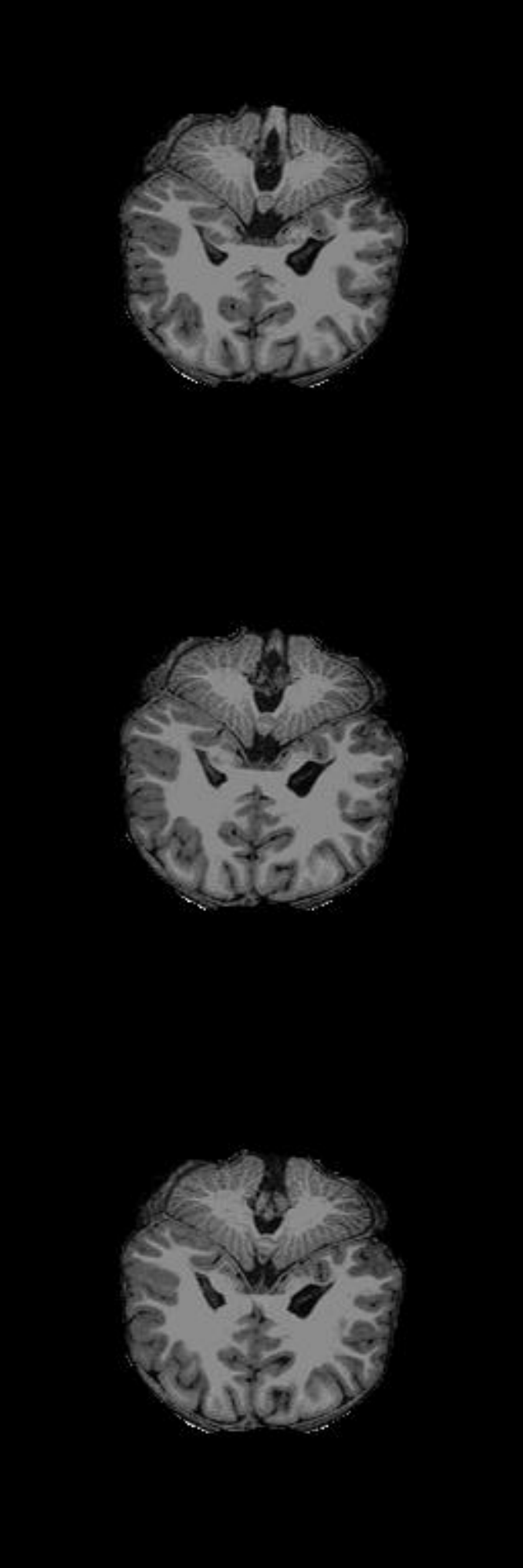}}%
\qquad
\subfigure[Ground truth]{%
\label{fig:2b}%
\includegraphics[height=3in, width=0.15\linewidth]{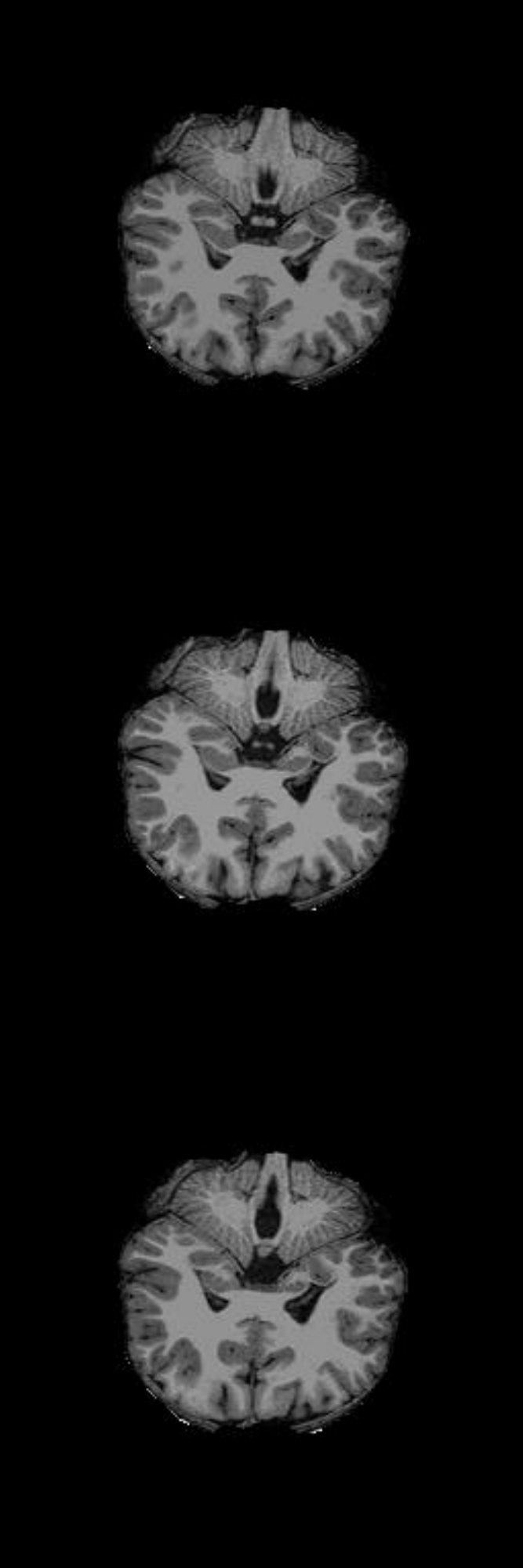}}%
\qquad
\subfigure[SAGAN33(47.52)]{%
\label{fig:3b}%
\includegraphics[height=3in, width=0.15\linewidth]{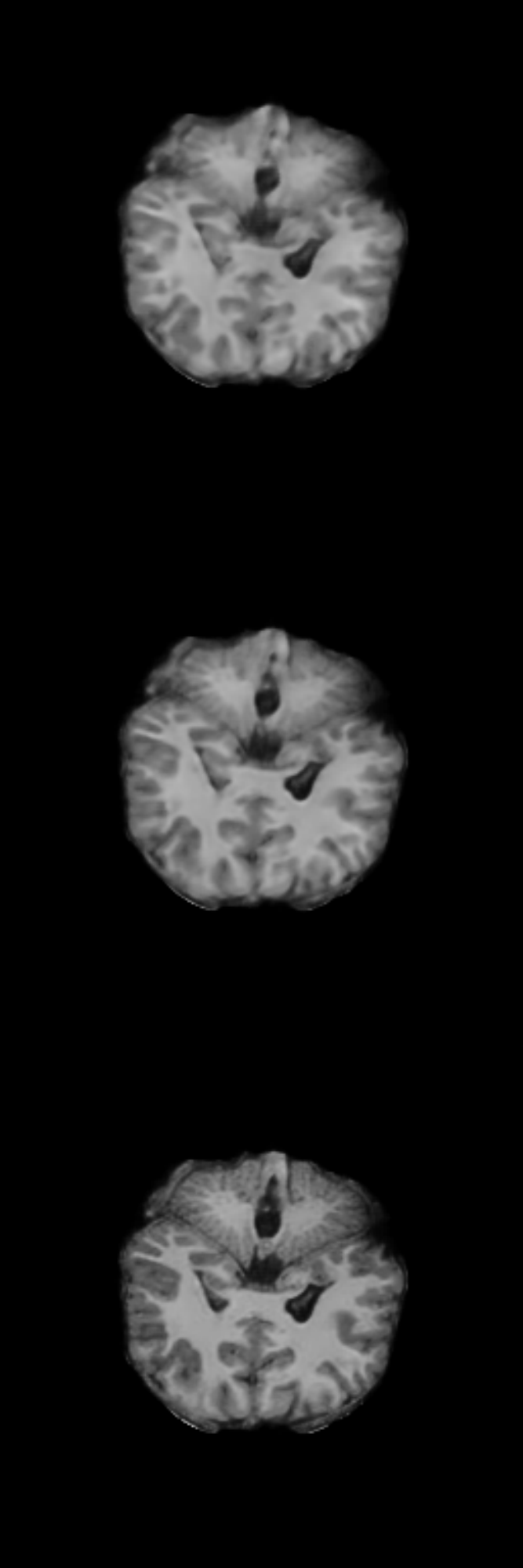}}
\caption{\rev{A sample of ASD Coronal slices reconstructed from the test set: (a) Input adjacent three slices; (b) Next adjacent three slices (ground truth);  (c) SAGAN33-reconstructed next adjacent three slices. PSNR (in dB) for ground truth vs respective model is given in brackets}}
\label{fig6}
\end{figure*}

\begin{figure*}%
\centering
\subfigure[Input]{%
\label{fig:1c}%
\includegraphics[height=3in, width=0.15\linewidth ]{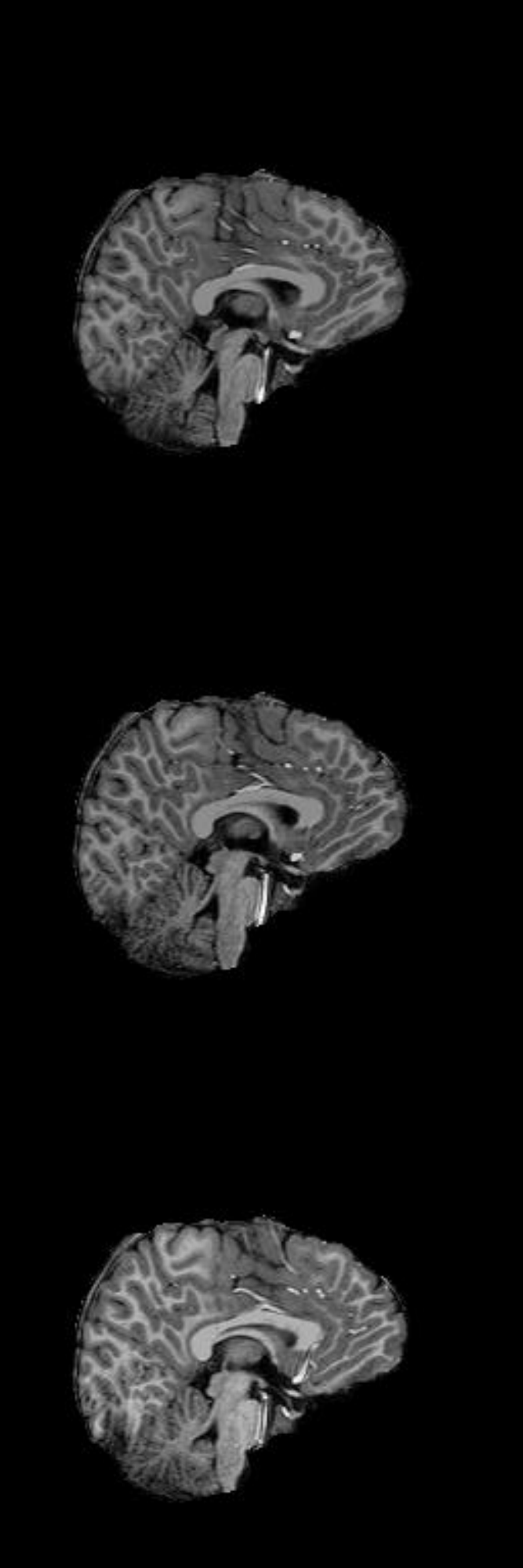}}%
\qquad
\subfigure[Ground truth]{%
\label{fig:2c}%
\includegraphics[height=3in, width=0.15\linewidth]{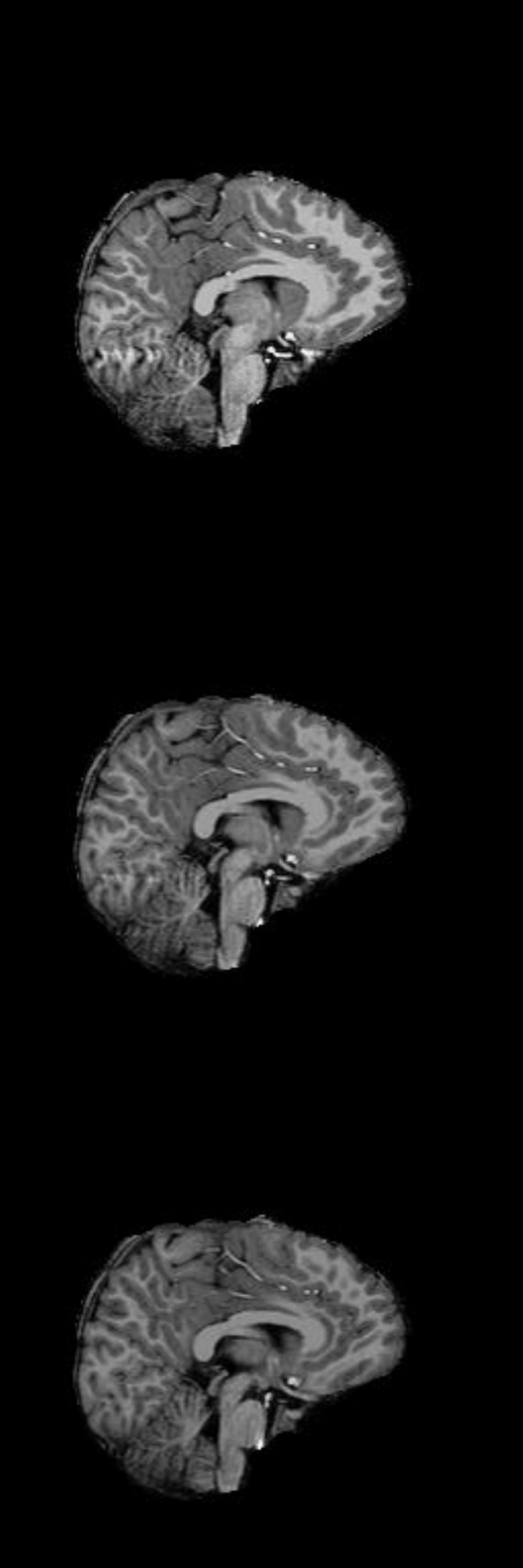}}%
\qquad
\subfigure[SAGAN33(46.95)]{%
\label{fig:3c}%
\includegraphics[height=3in, width=0.15\linewidth]{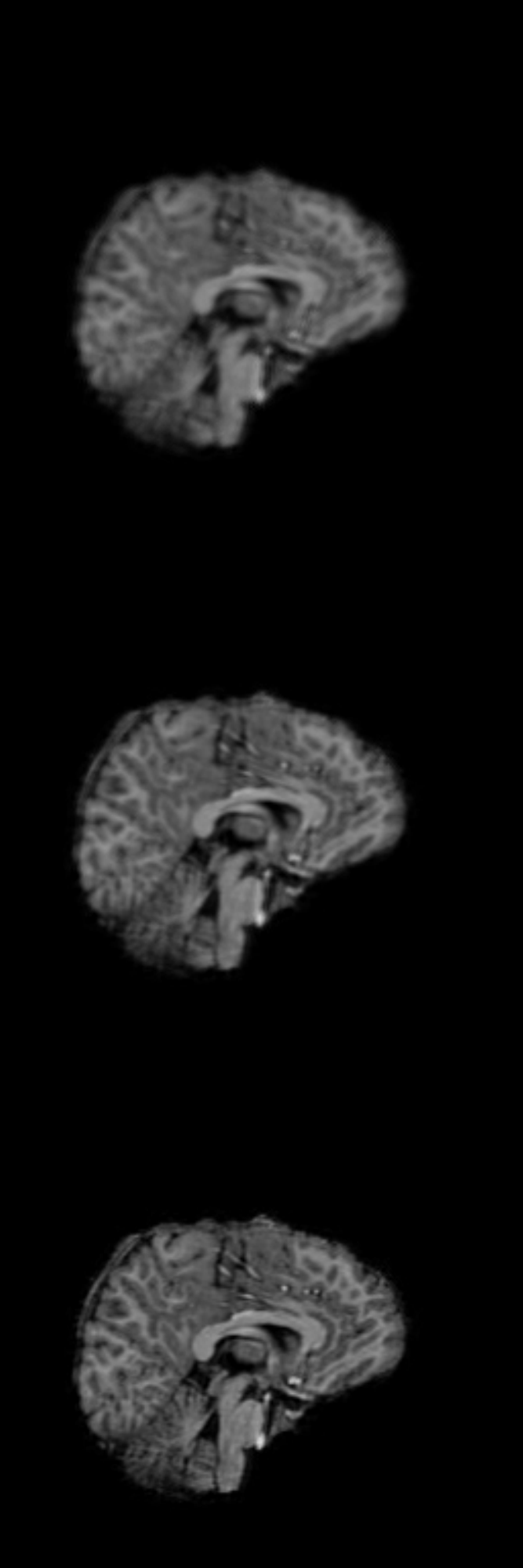}}
\caption{\rev{A sample of ASD Sagittal slices reconstructed from the test set: (a) Input adjacent three slices; (b) Next adjacent three slices (ground truth);  (c) SAGAN33-reconstructed next adjacent three slices. PSNR (in dB) for ground truth vs respective model is given in brackets}}
\label{fig7}
\end{figure*}

\subsection{Performance evaluation}
We assume significant sMRI structural differences between ASD and healthy subjects~\cite{asd49}, which should reflect via dissimilarities between reconstructed healthy and ASD slices. We examined the utility of the a) L2 and b) cosine loss functions for threshold-based outlier detection. The L2 loss is defined as in Eqn.~\eqref{eq:1}, and ranges from 0 to $\infty$. \rev{The cosine similarity loss or distance~\cite{asd67} computed for a pair of vectorized slice stacks $(X,Y)$ as shown in Eqn.~\eqref{eq:2}. The cosine loss enforces similarity between the generated and actual slices~\cite{asd68, asd85}, and ranges between 0 (for identical) to 1 (for highly dissimilar) slice stacks.} 
 
\begin{equation}
    Cos(X,Y) = 1-\frac{\sum^{m-1}_{i=0} \sum^{n-1}_{j=0} X_{i,j} \cdot Y_{i,j}}{\sqrt{ \sum^{m-1}_{i=0} \sum^{n-1}_{j=0} X_{i,j}^2} \sqrt{ \sum^{m-1}_{i=0} \sum^{n-1}_{j=0} Y_{i,j}^2} }
    \label{eq:2}
 \end{equation}

For classification, a threshold value $(\tau_{avg})$ is computed from the training samples as:
\begin{equation}
\tau_{avg} = \frac{1}{N} \left(\sum_{i=1}^N \frac{1}{n} \left(\sum_{j=1}^n  {Loss_{ij}} \right) \right)
\label{eq:3}
\end{equation}

\noindent where \textit{n} denotes the number of adjacent three-slice combinations per scan, \textit{Loss} is the reconstruction loss between predicted vs actual slices, and \textit{N} denotes the number of subjects in the training set. Test samples are classified based on the threshold value, \textit{i.e.}, if the reconstruction loss for the test sample is less than $\tau_{avg}$, it is marked as healthy or else as ASD. 
\rev{Two alternative thresholds are shown in Equations~\eqref{eq:3a} and~\eqref{eq:3b}. These are based on the maximum (or minimum) of the maximum (or minimum) reconstruction \textit{Loss} per subject.}
\begin{equation}
\tau_{max} = \max_{N} \left( \max_{n} (Loss_{ij}) \right)
\label{eq:3a}
\end{equation}
\begin{equation}
\tau_{min} = \min_{N} \left( \min_{n} (Loss_{ij}) \right)
\label{eq:3b}
\end{equation}

\rev{We use $\tau_{avg}$ as the threshold metric in our experiments, as this threshold reduces the number of false positives and false negatives, ensuring high sensitivity and specificity~\cite{asd66}. } For performance evaluation, we use model accuracy defined as:
\begin{equation}
  Accuracy \hspace{0.1cm}  = \frac{ TP + TN}{TP + TN + FP + FN} \times 100  
  \label{eq:4}
\end{equation}

\noindent where TP, TN, FP, and FN respectively denote the number of True Positives, False Positives, True Negatives and False Negatives. TP and TN represent correctly classified ASD and healthy samples, whereas FP and FN denote incorrect predictions.
We additionally report the area under the receiver operating characteristic curve (AUC) for evaluation.

\section{Results and discussion}\label{ExRes}
\textcolor{black}{SMRI-based ASD detection results obtained with the UNet33, GAN33 and SAGAN33 architectures on the Axial, Coronal and Sagittal slices are presented in this section. In all three architectures, adjacent 3-slices are input to reconstruct next three adjacent slices, and the reconstruction error is minimized during model training. }

\subsection{Slice reconstruction Quality}
Exemplar reconstructions achieved with UNet33, GAN33 and SAGAN33 for adjacent slices corresponding to the Axial modality are shown in Figures~\ref{fig4} and~\ref{fig5} respectively. Fig.~\ref{fig4} depicts a healthy test sample, while Fig.~\ref{fig5} presents an ASD sample. The first column in both figures depicts the input slices, and the second column presents the actual next-three slices. Columns 3--5 present reconstructions with the UNet33, GAN33 and SAGAN33, and parentheses values specify reconstruction Peak Signal-to-Noise Ratio (PSNR). 
In both cases, SAGAN33 achieves the highest PSNR and captures vivid details compared to UNet33 and GAN33. UNet33 trained with the L2 loss function reconstructs blurry images. GAN33 reconstructs images with good structural quality, while still performing inferior to SAGAN33. In UNet33 and GAN33, convolutions are limited to only the local domain of the convolution kernels, causing the network to overlook significant global structures. However, the SA mechanism in the SAGAN33 effectively captures global dependencies. \rev{Sample Coronal and Sagittal ASD slice reconstructions achieved by SAGAN33 are presented in Figures~\ref{fig6} and~\ref{fig7} respectively.} \rev{Evidently, the SAGAN33 adequately captures sMRI connectives. Ventricles of ASD subjects are larger and thicker than those of healthy subjects, as seen from Figures~\ref{fig4} and~\ref{fig5}. This observation is echoed by domain experts~\cite{asd46}. Clearly, visual cues in ventricular areas allow clinicians to distinguish ASD from healthy subjects, and likewise, can enable sMRI-based ASD diagnosis.} 

\subsection{ASD detection}
 Given a test sMRI slice-stack, the next three slices are reconstructed via the encoder-decoder networks described above, and the reconstruction loss compared against the threshold specified in Eqn.~\eqref{eq:3}. This threshold determines whether the sample is a healthy (inlier for which mean reconstruction loss $< \tau_{avg}$) or ASD (outlier, mean reconstruction loss $> \tau_{avg}$) sample, enabling unsupervised ASD detection. 
 
  \begin{table}[]
\caption{Accuracy and AUC scores for ASD detection.}
\centering
\begin{tabular}{cccccc}
\hline
\textbf{S.No} & \textbf{Model}  &\textbf{Objective} & \textbf{Distance}& \textbf{Accuracy} & \textbf{AUC} \\
& &\textbf{Metric} & \textbf{Metric}& \textbf{\%} & \\ \hline \hline
1  & UNet33  & L2 loss & {L2 loss} & 36.95 & 0.42 \\
\hline
2  & GAN33 & WGAN-GP & L2 loss  & 65.21 & 0.58 \\
&  & + 100L1 &  &  &  \\
\hline
3  & SAGAN33 & WGAN-GP & L2 loss & 80.43 & 0.60 \\
&  & + 100L1 &   &  &  \\
\hline
4  & SAGAN33 & WGAN-GP & Cosine  & 82.60 & 0.61 \\
&  & + 100L1 &   &  &  \\
\hline
5  & SAGAN33 & WGAN-GP & L2+Cosine  & 84.78 & 0.63 \\
&  & + 100L1 &   &  &  \\
\hline
6  & SAGAN33 & WGAN-GP & L2+  & \textbf{86.95} & \textbf{0.71} \\
&  & + 100L1 + & Cosine   &  &  \\
&  & Cosine &   &  &  \\
&  &  &   &  &  \\
\hline
\end{tabular}
\label{tab2}
\end{table}

\begin{table}[]
\caption{Confusion matrix values for SAGAN33 with different loss functions.}
\centering
\begin{tabular}{ccccccc}
\hline
\textbf{S.No} & \textbf{Obj Metric}  & \textbf{Dist Metric}  & \textbf{TP} & \textbf{TN} & \textbf{FP} & \textbf{FN} \\ \hline 
1 & WGAN-GP & L2 loss & 35 & 2  & 4  & 5  \\ 
& + 100 L1  & & & & &\\
\hline
2 & WGAN-GP& Cosine   & 36 & 2 & 4 & 4  \\ 
& + 100 L1 &   & & & & \\
\hline
3 & WGAN-GP & L2 loss + & 37  & 2 & 4 & 3 \\ 
& + 100 L1 & Cosine & & & & \\
& &  & & & & \\
\hline
4 & WGAN-GP & L2 loss +  & 37 & 3 & 3 & 3 \\ 
& + 100 L1 & Cosine & & & & \\
& + Cosine &  & & & & \\
&  &  & & & & \\
\hline
\end{tabular} 
\label{tab3}
\end{table}

We employed different metrics to train the encoder-decoder networks described above, and to compute the distance between the original and reconstructed test slices. Table~\ref{tab2} specifies the loss function employed for model training as the \textit{objective metric}, while the measure used to compute the distance between actual and reconstructed slices is specified as the \textit{distance metric}. The L2 and cosine distance metrics defined in Eqn.~\ref{eq:1} and~\ref{eq:2} were used for evaluating test samples. For model training, the UNet33 was trained with the L2 objective, while the GAN33 and SAGAN33 models were trained with the WGAN-GP+100L1 or WGAN-GP+100L1+Cosine loss objectives. 

Accuracy and AUC scores achieved by the UNet33, GAN33 and SAGAN33 networks with the different objective and distance metrics are listed in Table~\ref{tab2}. When the L2 distance metric is used for classification, the UNet33 model performs worst due to poor reconstruction quality to achieve an accuracy of 36.95\%. Without SA modules, the GAN33 network reconstructs slices with adequate structural detail and produces a fair accuracy of 65.21\%. The SAGAN33 which incorporates self-attention modules achieves the best reconstruction quality, and correspondingly the best ASD detection accuracy of 80.43\%. Overall, the SAGAN33 outperforms GAN33 by over 15\%. 

We also employed the cosine metric, and L2$+$Cosine measure as the distance metric with the SAGAN33 model. Table~\ref{tab2} confirms that the use of alternate distance metrics improves ASD detection accuracy. The cosine distance metric is more sensitive to outliers, and improves detection accuracy by over 2\%. A combination of the L2 and cosine metrics further improves detection performance, achieving an accuracy of 84.78\% and an AUC of 0.63. Finally, a SAGAN33 model incorporating cosine distance in the objective metric, along with the L2$+$cosine distance metric achieves the highest accuracy of 86.95\% and an AUC of 0.71. Also, while the small size of our dataset can make the models prone to overfitting, the GAN33 and SAGAN33 networks effectively address this issue via regularization applied in the objective. 

Table~\ref{tab3} extends the results in Table~\ref{tab2}, and presents the confusion matrix values for different objective--distance metric combinations employed with the SAGAN33 network. Given that the test set mainly comprised ASD samples (Table~\ref{tab1}), we note that the sensitivity or true-positive rate gradually increases as the distance metric changes from L2 to L2$+$cosine loss. The true-negative rate (or specificity) also increases slightly when the objective metric is modified to include the cosine loss. Cumulatively, these results convey that both the objective and distance metrics impact ASD detection sensitivity and specificity.   

We also note here that the longitudinal sMRI scans used in this study are heterogeneous, and were collected with different scanner settings (from different sites). Empirical results reveal that the proposed approach is robust to input data variations, and can be used in real-world situations where it is practically difficult to standardise scanning setups. 

\subsection{sMRI Imaging Modalities}
\begin{figure}[h]	
\centering
\includegraphics[width=8.7cm, height=6.5cm]{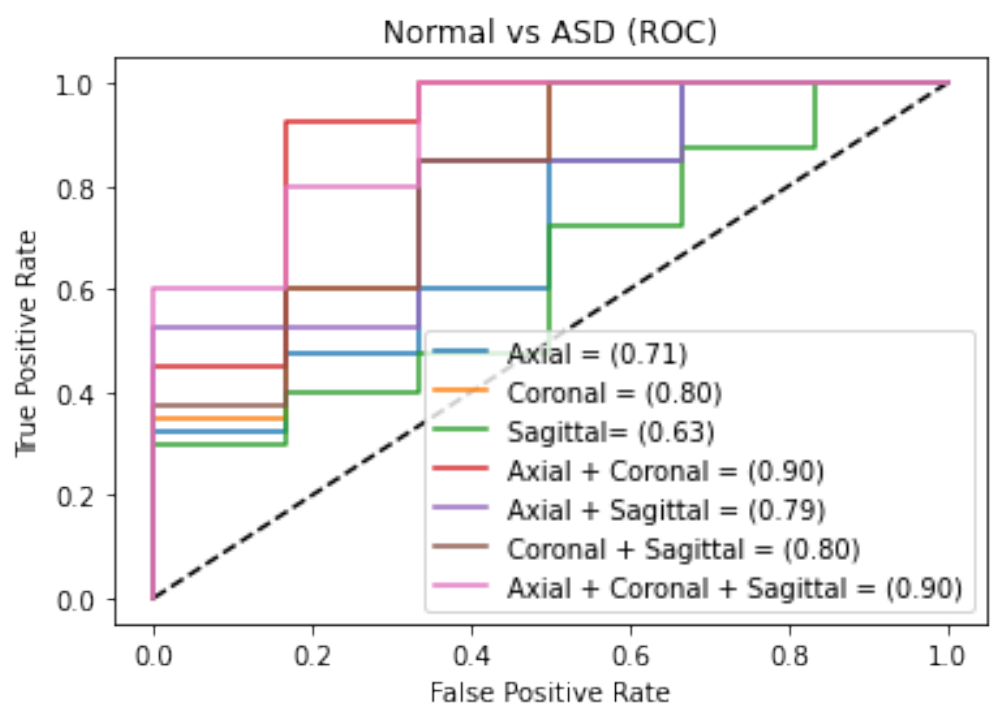}
\caption{\rev{ROC plot with AUC values illustrating SAGAN33 ASD detection performance on different sMRI planes.} \label {fig8}}
\end{figure}

To examine whether the detection performance is impacted by the sMRI imaging modality, the best performing model SAGAN33 was input with Axial, Coronal and Sagittal slices. Results are reported in Table~\ref{Planes}, \rev{and the corresponding Receiver Operating Curve (ROC) graph is plotted in Figure~\ref{fig8}}. The model objective was to minimize the WGAN-GP $+$ 100L1 $+$ Cosine loss, while the distance metric employed at test time was the L2 $+$ Cosine loss.  Among individual models, the Sagittal slices performed worst and Coronal slices best, achieving 20.4\% higher accuracy and 0.17 higher AUC over the Sagittal slices. Utilizing multimodal information for ASD detection was found to be more beneficial than unimodal slices. Higher detection accuracy was obtained on combining the Axial and Sagittal slices, and the best accuracy/AUC was achieved with a combination of the Axial and Coronal slices. Training the SAGAN33 with slices from all three imaging modalities however did not enhance the overall accuracy or AUC.

\begin{table}[h]
\centering
\caption{Performance of SAGAN33 on different planes of sMRI.}
\begin{tabular}{cccc}
\hline
{\textbf{S.No}} & \textbf{sMRI Modality} & \textbf{Accuracy} & \textbf{AUC}  \\
\hline \hline
1&Axial  & 86.95\%   & 0.71   \\ \hline
2&Coronal & {91.30\%}   & {0.80}   \\ \hline
3&Sagittal & 71.73\%   & 0.63   \\ \hline
4& Axial $+$ Coronal & \textbf{95.65\% }  &\textbf{ 0.90}  \\ \hline
5& Axial $+$ Sagittal & 89.13\%   & 0.79  \\ \hline
6& Coronal $+$ Sagittal & 91.30\%   & 0.80 \\ \hline
7& Axial $+$ Coronal $+$ Sagittal & \textbf{95.65\% }  & \textbf{0.90}  \\ \hline
\end{tabular}
\label{Planes}
\end{table}

\begin{figure}[h]%
\centering
\subfigure[Conv1-- Healthy subject]{%
\label{fig:9a}%
\includegraphics[height=1.7in, width=0.45\linewidth ]{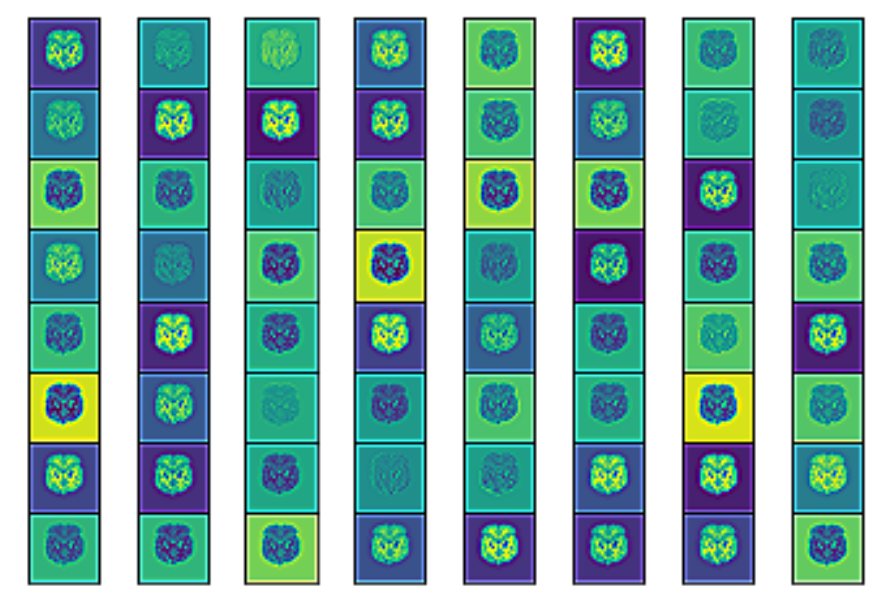}}%
\qquad
\subfigure[Conv1--ASD subject]{%
\label{fig:9b}%
\includegraphics[height=1.7in, width=0.45\linewidth]{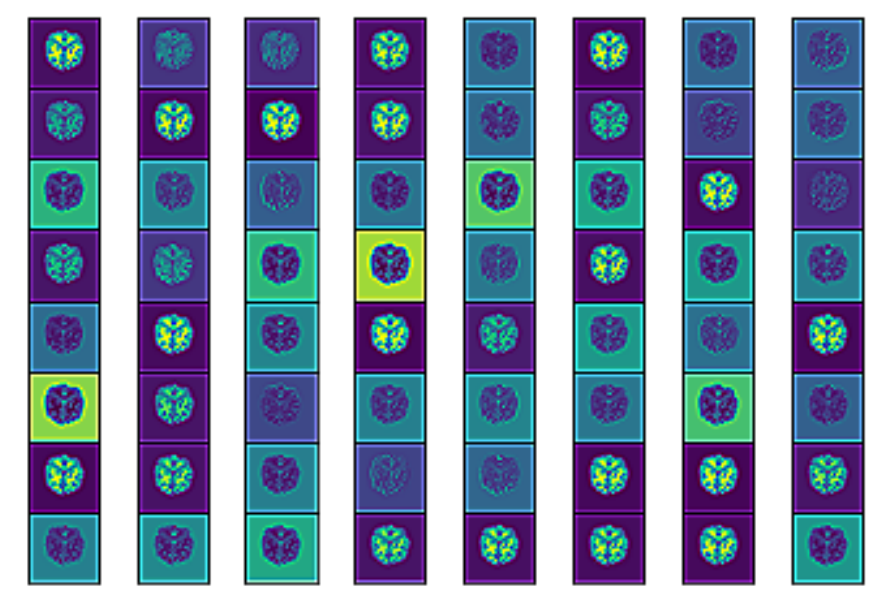}}
\qquad
\caption{\rev{Comparing Axial modality feature maps output by the first convolutional layer of SAGAN33 for healthy and ASD subjects.}}
\label{fig9}
\end{figure}

\subsection{GRAD-CAM Visualization}

\begin{figure}[h]%
\centering
\subfigure[Grad-CAM--Healthy subject]{%
\label{fig:10a}%
\includegraphics[height=1.5in, width=1.0\linewidth ]{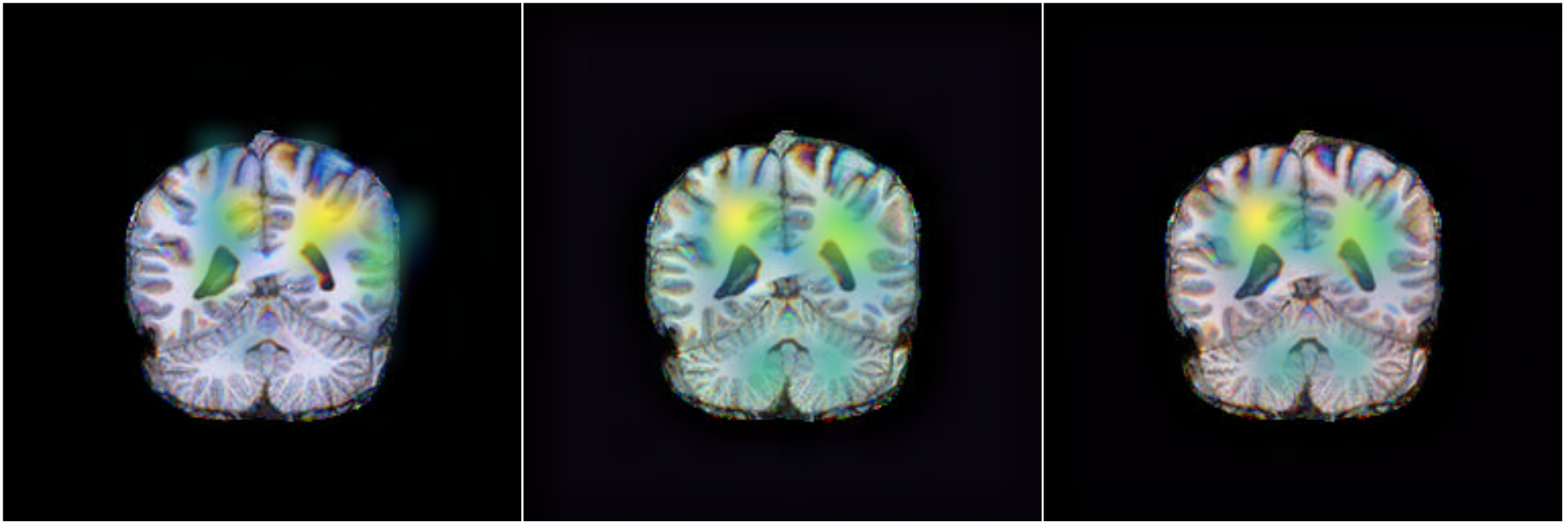}}%
\qquad
\subfigure[Grad-CAM--ASD subject]{%
\label{fig:10b}%
\includegraphics[height=1.5in, width=1.0\linewidth]{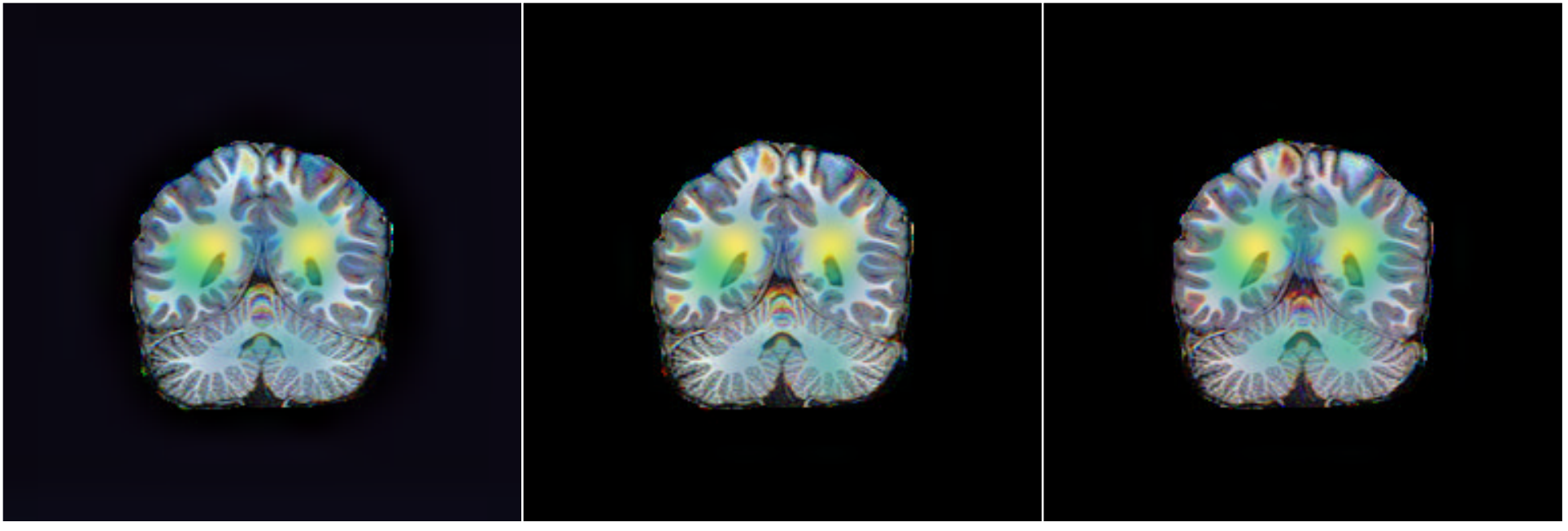}}%
\qquad
\caption{\rev{Grad-CAM visualizations for three adjacent Coronal slices from a healthy (top) and ASD (bottom) sample. Highlighted regions are found to be discriminative by the SA layers. }}
\label{fig10}
\end{figure}

\rev{To understand the features learned by the SAGAN model for encoding sMRI visual cues, we present two different visualization maps in Figures~\ref{fig9} and~\ref{fig10}. We used the Python package ELI5 to visualise Gradient-weighted class activation map (Grad-CAM)~\cite{asd80}. The Grad-CAM is used to create class-specific \textit{heatmap} visualizations in order to highlight the salient regions in the sMRI slices. Green or blue shades in the heatmap represent lesser importance, implying that the corresponding features are less significant from the model viewpoint, while the yellow, red and orange shades represent regions of moderate-to-high importance, implying that those features are \textit{attended to} by the model in order to model information or make inferences regarding the specific class.}

\rev{Figure.~\ref{fig9} visualizes outputs of the first SAGAN33 convolution layer for an exemplar healthy and ASD subject. Some visual differences can be noted from the class-specific heatmaps; given that the SAGAN33 is trained with healthy samples, very little attention can be noted on the ventricular regions which are key areas characterizing ASD subjects (see Figure.~\ref{fig5}). Visualizations from the fourth SA layer for three adjacent Coronal slices of a healthy vs ASD subject are presented in Figure \ref{fig10}. We can see that the model emphasizes on the hypothalamus, hippocampus, and amygdala, which are considered to be significant for ASD diagnosis~\cite{brainsci10070435}. The high-intensity regions reflect areas of interest to the model at prediction time.}   

\subsection{Comparison with other works}
\textcolor{black}{We compare our work with others which employ ABIDE I~\cite{asd41a} cross-sectional sMRI data to highlight the utility of longitudinal slices for ASD detection. Results are summarized in Table~\ref{Cross}.} \rev{Most baselines~\cite{asd61},~\cite{asd59},~\cite{asd58} and~\cite{asd69} employ brain region-specific features for model training. Outlier-based ASD detection similar to ours, via single sMRI slice reconstruction, is proposed in ~\cite{asd49}. Differently, we (a) utilized the longitudinal ABIDE-II data~\cite{asd41}, and (b) learned holistic structural connectives by reconstructing three-slice stacks in this study. Among baselines, a highest accuracy of 96.6\% is achieved by~\cite{asd49}, while our approach produces the second highest accuracy of 95.65\%. However, our model is trained with 20 times fewer data than in~\cite{asd49}; these results point to the effectiveness of employing multiple scans per subject acquired at different time-points for ASD diagnosis.} 

\begin{table}[h]
\centering
\caption{Comparison with cross-sectional ASD works.}
\begin{tabular}{ccccc}
\hline
\textbf{{S.No}} & \textbf{Methods} & \textbf{\# ASD}  & \textbf{\# Healthy} & \textbf{Accuracy}  \\
 &  & \textbf{subjects} & \textbf{subjects} & (\%)  \\
\hline \hline
1&\cite{asd61}   & 155  & 186 &61.69\%   \\ \hline
2&\cite{asd59}   & 78  & 104 &90.39\%   \\ \hline
3&\cite{asd49}   & 403  & 468 &\ \textbf{96.60\% }  \\ \hline
4&\cite{asd58}  & 518  & 567 &71.80\%   \\ \hline
5&\cite{asd69}  & 505  & 530 &82.00\%   \\ \hline
6& Proposed  & 20 & 15& {\textit{\textbf{95.65\%}}}   \\ \hline
\end{tabular}
\label{Cross}
\end{table}

\textcolor{black}{To motivate the utility of longitudinal sMRI data, we applied our best performing SAGAN33 model on cross-sectional sMRI scans corresponding to the Axial, Coronal and Sagittal planes. We randomly collected cross-sectional sMRI samples from ABIDE I~\cite{asd41a}, so that the total number of train and test samples equalled the longitudinal data size in this study. The results obtained are summarized in Table~\ref{Cross2}. When compared with Table~\ref{Planes}, rows 1-3, we see that detection accuracies with longitudinal data results are superior by 17-28\%, even if the accuracy/AUC trends are consistent for the different modalities. The Coronal modality performs best achieving an accuracy of 63\%, and an AUC of 0.64. Overall, these results support our rationale to perform ASD detection with longitudinal instead of cross-sectional data.} 

\begin{table}[h]
\centering
\caption{SAGAN performance on cross-sectional data from different planes.} 
\begin{tabular}{cccc}
\hline
\textbf{{S.No}} & \textbf{sMRI Modality} & \textbf{Accuracy} & \textbf{AUC } \\
\hline \hline
1&Axial  & 60.86\%   & 0.56   \\ \hline
2&Coronal & {63.04\%}   & {0.64}   \\ \hline
3&Sagittal & 54.34\%   & 0.52  \\ \hline
\end{tabular}
\label{Cross2}
\end{table}

\section{Conclusion and future work}\label{Con}
{We employ a GAN-based encoder-decoder framework on longitudinal sMRI slices, where the error between the reconstructed and actual adjacent three slice stacks is utilized to determine ASD samples as outliers. Three architectures, namely, the UNet, GAN and SAGAN were examined for reconstruction quality and therefrom, ASD detection performance. The SAGAN incorporating self-attention modules achieves the best reconstruction and detection accuracy, while the UNet trained with the L2 objective produces blurry reconstructions and the worst performance. Furthermore, both the \textit{objective metric} employed for model training and the \textit{distance metric} used for computing the reconstruction loss are found to significantly impact detection performance. The WGAN-GP+100L1 objective considerably improves performance of the GAN and SAGAN networks, while employing the cosine similarity instead of, or in combination with, the L2 norm as the distance metric also increases detection sensitivity.} 
Among other findings, of the three sMRI images planes-- Axial, Coronal and Sagittal, the Coronal mode yielded the highest accuracy, outperforming the Sagittal mode by around 20\%. This implies that sMRI structural connectivity is best encoded by Coronal information. Empirical results also revealed that the imaging modes are complementary; multimodal inputs improved accuracies over unimodal data by more than 5\%. Grad-CAM visualizations depicting \textit{regions-of-interest} to the network showed attention to the hypothalamus, hippocampus, and amygdala regions, considered important for ASD diagnosis~\cite{brainsci10070435}. Comparisons against ASD detection works examining cross-sectional data convey that longitudinal sMRI slices enable comparable performance with far fewer training data, and modeling structural brain connectivity with multiple scans over time per subject is beneficial.  
\textcolor{black}{Our unsupervised outlier detection framework would detect any deviation from the norm; apart from ASD, our framework could be extended to potentially tackle other disorders such as Attention deficit hyperactivity disorder (ADHD), Schizophrenia, \etc. Future work will focus on these extensions. \rev{Another interesting line of exploration would be to utilise multi-task learning for exploiting complementarities in (substantially available) cross-sectional and (sparse) longitudinal data for improving prediction accuracy as in~\cite{multitask2021}. We will also investigate architectures alternative to SAGAN such as Dense-Attentive GAN~\cite{asd84}.}}

\section*{Acknowledgment}
We thank Dr Leonardo Rundo for his valuable comments. 
\bibliography{journal_1}

\section*{Authors Biography}
\begin{wrapfigure}{l}{25mm} 
\includegraphics[width=1in,height=1.25in,clip,keepaspectratio]{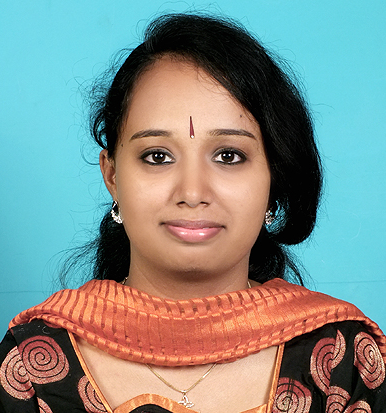}
\hspace{5mm}
\includegraphics[width=1in,height=1.25in,clip,keepaspectratio]{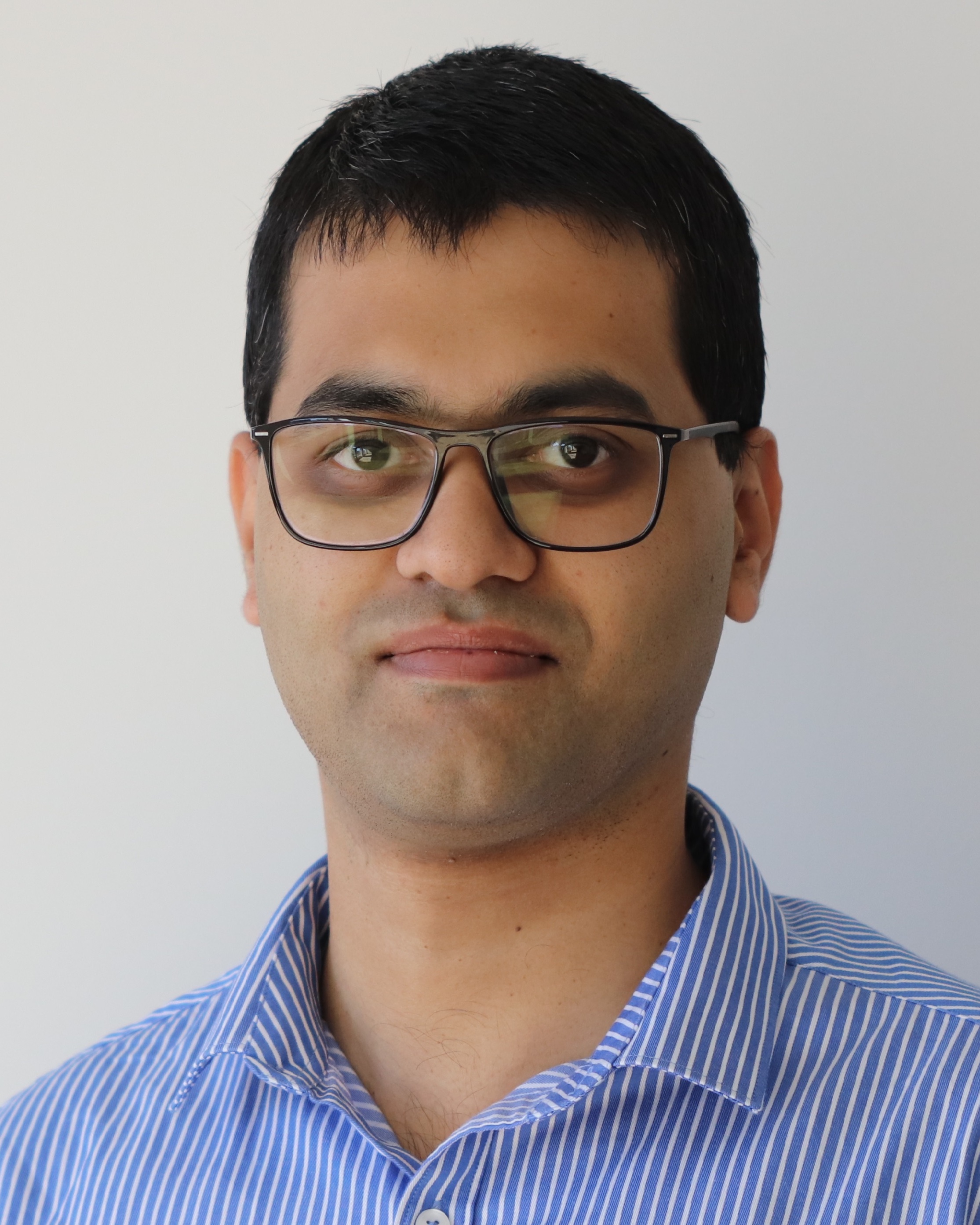}
\hspace{5mm}
\includegraphics[width=1in,height=1.25in,clip,keepaspectratio]{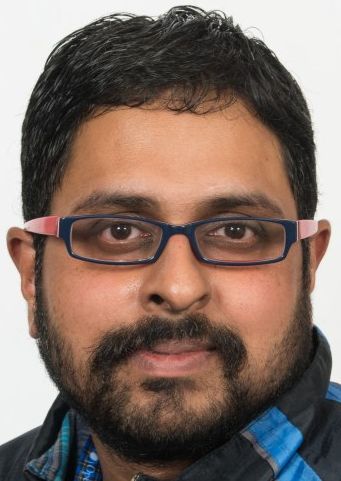}
\hspace{5mm}
\includegraphics[width=1in,height=1.25in,clip,keepaspectratio]{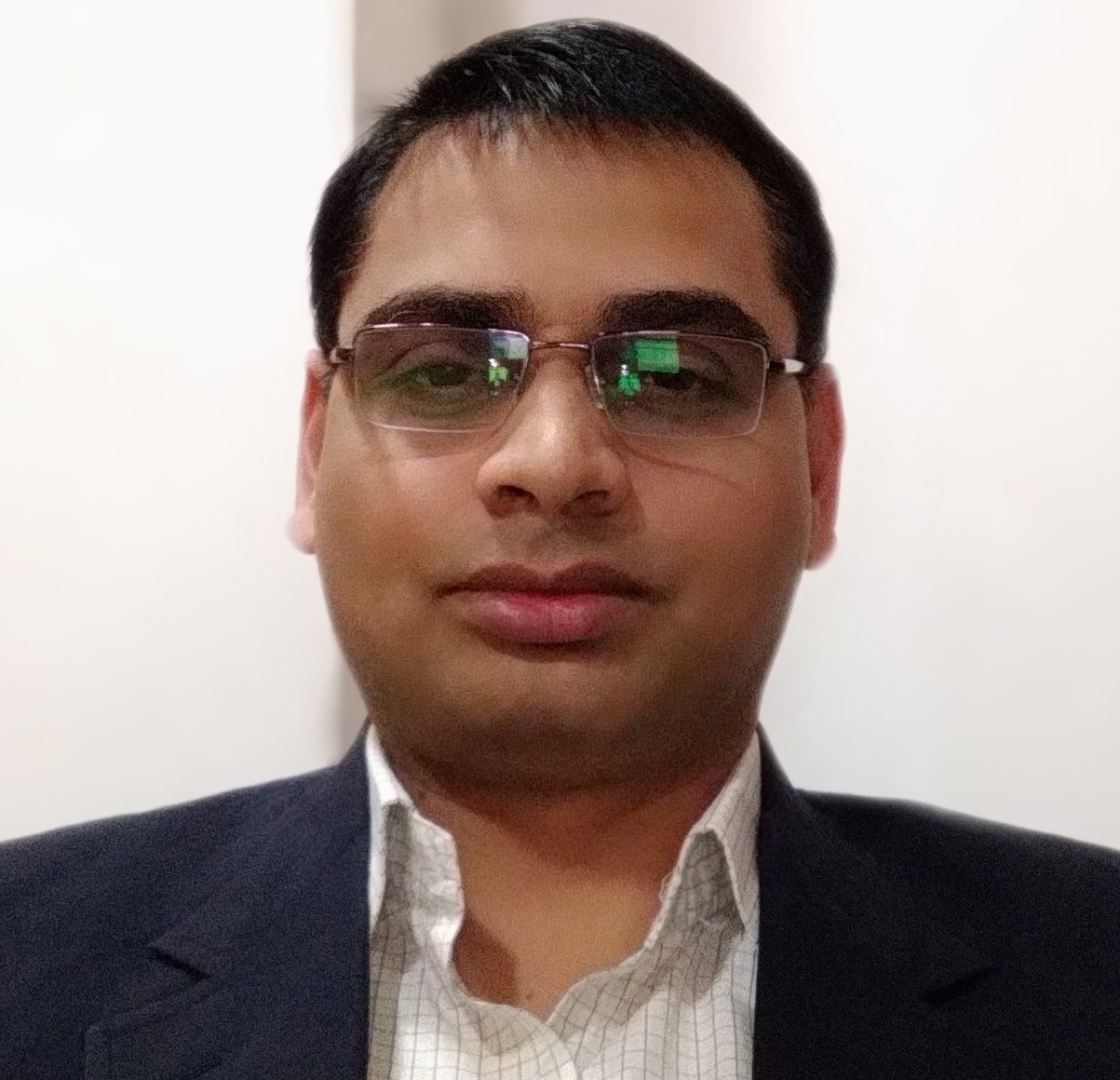}
\end{wrapfigure}\par
\textbf{Devika K} is a PhD student at Department of Electrical and Electronics Engineering, Amrita Vishwa Vidyapeetham, India. She completed her Masters in Computer Science at Amrita Vishwa Vidyapeetham, India in 2018. Her research interests include unsupervised and deep learning with applications in medical imaging. 
\\[11ex]
\textbf{Dwarikanath Mahapatra} received his Ph.D. in 2011 in the area of medical image analysis from the Department of Electrical and Computer Engineering at the National University of Singapore. He is a Senior Research Scientist at the Inception Institute of Artificial Intelligence, Abu Dhabi, UAE. His past affiliations include National University of Singapore, ETH Zurich Switzerland, IBM Research Australia. His research interests are in applying deep learning and machine learning to medical image analysis and computer vision. 
\\[6ex]
\textbf{Ramanathan Subramanian} received his Ph.D. in Electrical and Computer Engg. from NUS in 2008. He is an Associate Professor in University of Canberra, Australia. His past affiliations include IHPC (Singapore), U Glasgow (Singapore), IIIT Hyderabad (India), IIT Ropar (India) and UIUC-ADSC (Singapore). His research focuses on Human-centered computing, Interactive analytics and Explainable machine learning. He is an IEEE Senior Member and a member of the ACM and AAAC.
\\[5ex]
\textbf{V Ramana Murthy Oruganti}  received his Masters and PhD degrees in Electrical Engineering from IIT Delhi, India. He is an Assistant Professor in Department of Electrical and Electronics Engineering, Amrita Vishwa Vidyapeetham, India. His past affiliations include NUS (Singapore), NTU (Singapore), University of Canberra (Australia) and Carnegie Mellon University (US). His research focuses on medical image processing and affective computing. He is a Member of IEEE and the ACM.\par
  
\end{document}